\documentclass[%
 reprint,
nofootinbib,
 amsmath,amssymb,
 aps,
]{revtex4-1}
\usepackage{color}
\usepackage{graphicx}
\usepackage{dcolumn}
\usepackage{bm}
\usepackage{hyperref}
\usepackage{longtable}
\usepackage{supertabular}

\begin{document}

\preprint{APS/123-QED}

\title{ Vibrational and rotational excited states within Bohr Hamiltonian with deformation-dependent mass formalism}

\author{M. Chabab}
\author{A. Lahbas}%
 \author{M. Oulne}
  \email{corresponding author: oulne@uca.ma}

\affiliation{%
High Energy Physics and Astrophysics Laboratory, Department of Physics, \\ Faculty of Sciences Semlalia, Cadi Ayyad University,\\
P. O. B. 2390, Marrakesh 40000, Morocco\\
}%


\date{\today}

\begin{abstract}
In a recent work (Phys.Rev.C84, 044321, 2011) M.J. Ermamatov and P.R. Fraser have studied   rotational and vibrational excited states of axially symmetric nuclei within the Bohr Hamiltonian with  different mass parameters. However, the energy formula that the authors have used contains some     inaccuracies. So, the numerical results they obtained seem to be controversial. In this paper, we revisit all calculations related to this problem and determine the appropriate formula for the energy spectrum.   Moreover, in order to improve such calculations, we reconsider this problem within the  framework of    the deformation dependent mass formalism. Also, unlike the work of Bonatsos et al. (Phys.Rev.C83, 044321, 2011) in which the mass parameter has not been considered, we will show the importance of this parameter and its effect on numerical predictions.
\begin{description}
\item[PACS numbers] 21.10.Re, 21.60.Ev, 23.20.Lv, 27.70.+q
\end{description}
\end{abstract}

\pacs{Valid PACS appear here} 
\maketitle


\section{\label{sec:level1}Introduction}
Thanks to its relatively simple structure the Bohr Hamiltonian ~\cite{bohr} continues to play an undeniable role in the study of nuclear structure within collective models ~\cite{bohr2,eisenberg} in competition with more sophisticated methods such as
Quasiparticle Random Phase Approximation (QRPA) \cite{rin80,qrpa11} and
Interacting Boson Model (IBM) \cite{iachello87}. Also, its advantage in respect to these microscopic methods resides in its ability to provide collective states eigenenergies and corresponding wave functions of nuclei in analytical form. So far, the Bohr Hamiltonian has been widely used with a constant mass parameter \cite{fort03,fort0430,fot04,fortunato,bonat04,bonat07,bonat11}. Recently, this assumption has been reexamined in the framework of Deformation Dependent Mass Formalism (DDMF) \cite{bonatsos2,bona2013} emphasizing the mass tensor of the collective Hamiltonian cannot be taken as a constant but it has to depend on the collective coordinates. Such a formalism allows enhancing the precision of numerical calculations of nuclear characteristics. Moreover, Jolos et al. \cite{jolos0,jolos,jolos2,jolos3} have shown this mass parameter should  split into ground state band, $\beta$-band and $\gamma$-band coefficients for deformed nuclei. Each coefficient is set to its average value over the wave function of the corresponding band state. Following this later procedure, M.J. Ermamatov et al. have studied rotational and vibrational spectra of axially symmetric nuclei \cite{ermamatov}. Their calculations have been based on an analytical energy formula that the authors claimed they have obtained in a previous work \cite{ermamatov2}.
However, the used formula in  \cite{ermamatov} together with the corresponding wave functions were inaccurate as we will show in this paper.
 Therefore, the calculated transition rates by the same authors are also questionable. Besides, the Bohr Hamiltonian's dependence on two separable collective coordinates $\beta$ and $\gamma$ where $\beta$ also represents  nuclear shape deformation enables one to choose nuclear collective potentials as a sum of two separate terms, namely: a $\beta$-potential $V(\beta)$ and a $\gamma$-term $V(\gamma)$. In the present paper where we revisit the M.J. Ermamatov et al. work \cite{ermamatov} with the purpose to improve their calculations  within DDMF, the potential term $V(\beta)$ is chosen to be equal to Davidson potential \cite{davidson} as in \cite{ermamatov} and the $\gamma$-potential $V(\gamma)$ taken to be equal to the harmonic oscillator. Such a problem  has been solved in \cite{bonatsos2}  but with equal mass coefficients by means of  supersymmetric quantum mechanical method (SUSYQM) \cite{Cooper1,Cooper2}. Furthermore, we will display the essential role played by the mass parameter in the evaluation of nuclear characteristics unlike the Bonatsos et al. work \cite{bonat04,bonat07,bonat11,bonatsos2,bona2013} in which this parameter has been hidden. Thus, the eigenenrgies formula and the corresponding wave functions are derived by means of the asymptotic iteration method (AIM) \cite{Iam}. This method has proved to be a useful tool when dealing with physical problems involving Schr\"{o}dinger type equations \cite{Iam1,Iam2,Iam3}.\\
    This paper is organized as follows : In Section II the position-dependent mass formalism is briefly described. In section III, we propose the Bohr Hamiltonian with three different mass coefficients, that we use in Section IV in accordance with deformation-dependent mass formalism. The exact separation of the Bohr hamiltonian in the case of axially symmetric prolate deformed nuclei and the solutions of angular equation are achieved in section V. The radial equation is given in Section VI. Analytical expressions for the energy levels and excited-state wave functions are presented in Sections VII and VIII respectively, while the $B(E2)$ transition probabilities are given in the Section IX. Finally, the section X is devoted to the numerical calculations for energy spectra and $B(E2)$ transition probabilities with their comparisons with experimental data and the available IBM ones, while Section XI contains the conclusion. An overview of the asymptotic iteration method is given in Appendix A. While in  Appendix B, we give the used formulas  for the calculations of  $B(E2)$.
 \section{\label{sec:level1}  POSITION-DEPENDENT MASS FORMALISM}
The general form of the Hamiltonian with effective mass depending on position has been originally introduced by  Von Roos ~\cite{roos83},
\begin{multline}
\tiny H=-\frac{\hbar ^2}{4}$\Big[$ m^{\delta'}(x)\nabla m^{\kappa'}\nabla m^{\lambda'}\\ +m^{\lambda'}(x)\nabla m^{\kappa'}\nabla m^{\delta'}$\Big]$+V(x),
  \label{eq1}
\end{multline}
where $V$ is the relevant potential and the parameters $\delta', \kappa', \lambda'$ are constrained by the condition $\delta'+\kappa'+ \lambda' = -1$. Assuming a position dependent mass of the form ~\cite{quesne2004}

\begin{multline}
  m(x)=m_0M(x), M(x)=\frac{1}{(f(x))^2}, f(x)=1+g(x),
  \label{eq2}
\end{multline}
where $m_0$ is a constant mass and $M(x)$ is a dimensionless position-dependent mass, the Hamiltonian (\ref{eq1}) becomes  ~\cite{quesne2004}
\begin{multline}
    H=-\frac{\hbar ^2}{4m_0}$\Big[$ f^{\delta}(x)\nabla f^{\kappa}\nabla f^{\lambda}\\+f^{\lambda}(x)\nabla f^{\kappa}\nabla f^{\delta}$\Big]$+V(x),
  \label{eq3}
\end{multline}
with $\delta+\kappa+ \lambda = 2$. It is known ~\cite{quesne2004} that this Hamiltonian can be put into the form
\begin{equation}
H=-\frac{\hbar^2}{2m_0}\sqrt{f(x)}\nabla f(x)\nabla \sqrt{f(x)}+V_{eff}(x),   \label{eq4}
\end{equation}
with
\begin{multline}
V_{eff}(x)=V(x)+\frac{\hbar^2}{2m_0}$\Big[$\frac{1}{2}(1-\delta-\lambda)f(x)\nabla^2f(x)\\
+\big(\frac{1}{2}-\delta\big)\big(\frac{1}{2}-\lambda\big)(\nabla f(x))^{2}$\Big],$
    \label{eq5}
\end{multline}
where $\delta$ and $\lambda$ are free parameters.

\section{\label{sec:level1} Bohr Hamiltonian with MASS COEFFICIENTS }
In the laboratory frame, the Bohr Hamiltonian can be written as~\cite{jolos}
\begin{equation}
    H=\frac{1}{4}\left(\sum_ {\mu}\pi^+_{2\mu}\pi_{2\mu}\frac{1}{B(\alpha_2)}+\frac{1}{B(\alpha_2)}\sum_ {\mu}\pi^+_{2\mu}\pi_{2\mu}\right)+V(\alpha_2)
  \label{eqq6}
\end{equation}
Where $\alpha_{2\mu}$ is a collective variable and $\pi_{2\mu}$ is an operator of the conjugate momentum. In the intrinsic frame we obtain from Eq. (\ref{eqq6})
\begin{multline}
    H=-\frac{\hbar^2}{4B(\beta,\gamma)}$\Bigg($\frac{1}{\beta^{4}}\frac{\partial}{\partial\beta} {\beta^4}\frac{\partial}{\partial\beta}+ \frac{1}{\beta^2\sin3\gamma}\frac{\partial}{\partial\gamma}\tiny{\sin3}\gamma\frac{\partial}{\partial\gamma}\\-
   \frac{1}{4\beta^2}\sum_ {k=1,2,3}\frac{Q_{k}^{2}}{\sin^2(\gamma-\frac{2}{3}\pi k)}$ \Bigg)$-$\Bigg($\frac{1}{\beta^4}\frac{\partial}{\partial\beta} {\beta^4}\frac{\partial}{\partial\beta}\\+ \frac{1}{\beta^2\sin3\gamma}\frac{\partial}{\partial\gamma}\tiny{\sin3}\gamma\frac{\partial}{\partial\gamma}-
   \frac{1}{4\beta^2}\sum_ {k=1,2,3}\frac{Q_{k}^{2}}{\sin^2(\gamma-\frac{2}{3}\pi k)}$ \Bigg)$\\\times\frac{\hbar^2}{4B(\beta,\gamma)}+V(\beta,\gamma)
  \label{eq7}
\end{multline}
For small amplitudes of $\gamma$-vibration around $\gamma=0$ and $\beta$-vibration around $\beta=\beta_0\neq0$, the collective coordinates could be considered as separable in the axial symmetry nuclei case. Thus, we can consider three separable states of nuclei, namely : the ground state, the $\beta$ and $\gamma$ vibrational states. Each one of these states will have its own mass parameter equal to its average value over the wave function of the state under consideration :

\begin{enumerate}
\item The ground state mass parameter
\begin{equation} \label{eq8}
   \langle g.s.\mid B(\beta,\gamma) \mid g.s. \rangle \equiv B_{rot}
    \end{equation}
     where we consider the ground state rotational band;
\item   the $\gamma$-mass parameter
 \begin{equation} \label{eq9}
   \langle \gamma \mid B(\beta,\gamma) \mid \gamma \rangle \equiv B_{\gamma}
\end{equation}
where we consider $\gamma$-vibrational state;
\item   The $\beta$-mass parameter
\begin{equation} \label{eq10}
   \langle \beta \mid B(\beta,\gamma) \mid \beta \rangle \equiv B_{\beta}
     \end{equation}
where we consider $\beta$-vibrational state.
\end{enumerate}
The procedure described above assumes the use of projection operators.
Using Eqs. (\ref{eq8}-\ref{eq10}), we obtain from Eq. (\ref{eq7}) the
following  Hamiltonian
\begin{multline}
    H=-\frac{\hbar^2}{2 \langle i\mid B\mid  i \rangle}$\Bigg($\frac{1}{\beta^4}\frac{\partial}{\partial\beta} {\beta^4}\frac{\partial}{\partial\beta}+ \frac{1}{\beta^2\sin3\gamma}\frac{\partial}{\partial\gamma}\tiny{\sin3}\gamma\frac{\partial}{\partial\gamma}\\-
   \frac{1}{4\beta^2}\sum_ {k=1,2,3}\frac{Q_{k}^{2}}{\sin^2(\gamma-\frac{2}{3}\pi k)}$ \Bigg)$+V(\beta,\gamma)
  \label{eq11}
\end{multline}
where $i=g.s.,$ $\beta$ or $\gamma $ band depending on which state is considered.
In the case of a small axially symmetric deformation of nuclei, the Bohr Hamiltonian with three different mass coefficients can be written as~\cite{jolos}
\begin{equation}
H=H_{rot}+H_{\gamma}+H_{\beta}
 \label{eq12}
\end{equation}
where
\begin{equation}
H_{rot}=\frac{\hbar^2}{6B_{rot}\beta^2}(\hat{\vec{Q}}^2-\hat{Q}_3^2)
\label{eq13}
\end{equation}
\begin{equation}
H_{\gamma}=-\frac{\hbar^2}{2B_{\gamma}\beta^2}\frac{1}{\gamma}\frac{\partial}{\partial\gamma}\gamma\frac{\partial}{\partial\gamma}+\frac{\hbar^2}{2B_{\gamma}}\frac{\hat{Q}_3^2}{4\beta^2\gamma^2}+\frac{V(\gamma)}{\beta^2}
\label{eq14}
\end{equation}
and
\begin{equation}
H_{\beta}=-\frac{\hbar^2}{2}\left(\frac{1}{B_{\beta}}\frac{\partial^2}{\partial\beta^2}+\frac{2}{B_{\gamma}}\frac{1}{\beta}\frac{\partial}{\partial\beta}+\frac{2}{B_{\beta}}\frac{1}{\beta}\frac{\partial}{\partial\beta}\right)+V(\beta)
 \label{eq15}
\end{equation}
\section{\label{sec:level1} Bohr Hamiltonian with different deformation-dependent mass parameters}
To construct a Bohr Hamiltonian with a mass depending on the deformation coordinate $\beta$, in accordance with the formalism described in section II,
\begin{equation}\label{eq16}
B=\frac{ \langle i\mid B_0\mid  i \rangle}{(f(\beta))^2}
\end{equation}
we have to follow the procedure in ~\cite{jolos}.
Since the deformation function $f$ depends only on the radial coordinate $\beta$, only the $\beta$ part of the resulting equation will be affected.

The final result reads ~\cite{bonatsos2}
\begin{multline}
 \frac{\hbar^2}{2\langle i\mid B_0\mid  i \rangle}  $\Big($ -\frac{\sqrt{f}}{\beta^4}\frac{\partial}{\partial\beta} {\beta^4f}\frac{\partial}{\partial\beta}\sqrt{f}- \frac{f^2}{\beta^2\sin3\gamma}\frac{\partial}{\partial\gamma}\sin3\gamma\frac{\partial}{\partial\gamma}\\+
  \frac{f^2}{4\beta^2}\sum_ {k=1,2,3}\frac{Q_{k}^{2}}{\sin^2(\gamma-\frac{2}{3}\pi k)} $\Big)$\Psi+V_{eff}\Psi=E\Psi  \label{eq17}
\end{multline}
with,
\begin{multline}
V_{eff}=V(\beta,\gamma)+\frac{\hbar^2}{2\langle i\mid B_0\mid  i \rangle}  $\Big($ \frac{1}{2}(1-\delta-\lambda)f\bigtriangledown^2f\\
+(\frac{1}{2}-\delta)(\frac{1}{2}-\lambda)(\bigtriangledown f)^{2} $\Big)$
    \label{eq18}
\end{multline}
\section{\label{sec:level1} Separation of the bohr hamiltonian or Axially symmetric prolate deformed nuclei}
Exact separation of the variables $\beta$ and $\gamma$ may be
achieved when the potential is chosen as in ~\cite{wilet,fortunato}:
\begin{equation}
V(\beta,\gamma)=U(\beta)+\frac{f^2}{\beta^2}W(\gamma)
    \label{eq19}
\end{equation}
where the potential $W(\gamma)$ has a minimum around  $\gamma=0$. Then, one can write the angular momentum of Eq.(\ref{eq17}) in the form~\cite{iachello}
\begin{multline}
\sum_ {k=1,2,3}\frac{Q_{k}^{2}}{\sin^2(\gamma-\frac{2}{3}\pi k)} \\ \approx \frac{4}{3}(Q_1^2+Q_2^2+Q_3^2)+Q_3^2\left(\frac{1}{\sin^2\gamma}-\frac{4}{3}\right)
    \label{eq20}
\end{multline}
In the same context, we consider a wave function of the form~\cite{iachello}
\begin{equation}
\Psi(\beta,\gamma,\theta_i)=F_{n_{\beta}L}(\beta)\eta_{n_{\gamma},K}(\gamma)\mathcal{D}_{M,K}^L(\theta_i)
 \label{eq21}
\end{equation}
where $\mathcal{D}(\theta_i)$ are Wigner functions of the Euler angles $\theta_i(i=1,2,3)$,  and $L$ is the total angular momentum, where $M$ and $K$ are the eigenvalues of the projections of angular momentum on the laboratory-fixed $z$-axis and the body-fixed $z'$-axis, respectively.
As a result, Eq. (\ref{eq17})  can be approximately separated into three equations:
\begin{multline}
   \Bigg[\frac{\hbar^2}{2\langle i\mid B_0\mid  i\rangle}\Big( -\frac{\sqrt{f}}{\beta^4}\frac{\partial}{\partial\beta} {\beta^4f}\frac{\partial}{\partial\beta}\sqrt{f}+\frac{f^2}{\beta^2}\Lambda\\+\frac{1}{2}(1-\delta-\lambda)f\bigtriangledown^2f
+(\frac{1}{2}-\delta)(\frac{1}{2}-\lambda)(\bigtriangledown f)^{2}\Big)\\+V(\beta)\Bigg]F_{n_{\beta}L}(\beta)=E F_{n_{\beta}L}(\beta)  \label{eq22}
\end{multline}

\begin{multline}
   \Bigg[-\frac{\hbar^2}{2B_{\gamma}}\Bigg(\frac{1}{\sin3\gamma}\frac{\partial}{\partial\gamma}\sin3\gamma\frac{\partial}{\partial\gamma}-
  \frac{K^2}{4}\frac{1}{\sin^2\gamma}\Bigg)\\+W(\gamma) \Bigg]\eta_{n_{\gamma},K}(\gamma)=\bar\Lambda\eta_{n_{\gamma},K} (\gamma)\label{eq23}
\end{multline}
and
\begin{equation}
  \frac{ \hbar^2}{6B_{rot}}\Big(\hat{\vec{Q}}^2-Q_3^2\Big)\mathcal{D}_{M,K}^L(\theta_i)=\Lambda' \mathcal{D}_{M,K}^L(\theta_i)\label{eq24}
\end{equation}
The eigenvalues of the rotational part equation (\ref{eq24}) are easily obtained since $\hat{\vec{Q}}^2$ is the quadratic casimir operator of $O(3)$ and  $\hat{\vec{Q}}_3^2$ is the projection of the angular momentum on the $z$-axis,
\begin{equation}
 \Lambda'= \frac{ \hbar^2}{6B_{rot}}\Big(L(L+1)-K^2\Big)
 \label{eq25}
\end{equation}
Note that Eq. (\ref{eq23}) for $\gamma \approx 0$ can be treated as in Ref.~\cite{iachello}.

For the $\gamma$-part, we use a harmonic oscillator potential \cite{ermamatov}
\begin{equation}
W(\gamma)=\frac{1}{2}(\beta_0^4C_{\gamma})\gamma^2
 \label{eq26}
\end{equation}
where $\beta_0$ denotes the position of the minimum of the potential in $\beta$ and $C_{\gamma}$ is a free parameter. In this case, Eq. (\ref{eq23}) transforms
into the usual harmonic oscillator equation
\begin{multline}
   \Bigg[-\frac{\hbar^2}{2B_{\gamma}}\Bigg(\frac{1}{\gamma}\frac{\partial}{\partial\gamma}\gamma\frac{\partial}{\partial\gamma}-
  \frac{K^2}{4}\frac{1}{\gamma^2}\Bigg)\\+\frac{1}{2}(\beta_0^4C_{\gamma})\gamma^2 \Bigg]\eta_{n_{\gamma},K}(\gamma)=\bar\Lambda\eta_{n_{\gamma},K}(\gamma),\label{eq27}
\end{multline}
To solve this equation through AIM, we propose the following ansatz for the $\gamma$-part  eigenvectors $\eta_{n_{\gamma},K}(\gamma)$,
\begin{align}
\eta_{n_{\gamma},K}(\gamma)=\gamma^{|K/2|}e^{-\frac{\gamma^2}{2g}}\Gamma_{n_{\gamma},K}(\gamma)
\label{eq281}
\end{align}
with $g =\frac{1}{\beta_0^2}\frac{\hbar}{\sqrt{B_{\gamma}C_{\gamma}}}$. For this form of the angular wave function, the $\gamma$-part equation (\ref{eq27}) reduces to a standard form given in the Appendix Eq. (\ref{A.1}). According to the AIM procedure, the eigenvalues are calculated by means of the termination condition Eq. (\ref{A.3}) and the recurrence relations Eq. (\ref{A.4}), hence one can derive the generalized form of the eigenvalues,
\begin{align}
&&&&\bar\Lambda=\frac{2}{g}\frac{\hbar^2}{B_{\gamma}}\Big(2\tilde n_{\gamma}+\frac{K}{2}+1\Big), && \tilde n_{\gamma}=0,1,2,...,
\label{eq281}
\end{align}
 By inserting $ \tilde n_{\gamma}=\frac{n_{\gamma}-|K/2|}{2}$ in Eq. (\ref{eq281}), where $n_{\gamma}$ is the quantum number related to $\gamma$-oscillations,  one obtains
\begin{align}
&&&&\bar\Lambda=\frac{2}{g}\frac{\hbar^2}{B_{\gamma}}\big(n_{\gamma}+1\big), && n_{\gamma}=0,1,2,...,
\label{eq28}
\end{align}
As a result, we found,
\begin{equation}
\frac{B_{\beta}}{\hbar^2}\Lambda=\Bigg(\frac{2}{g}\frac{B_{\beta}}{B_{\gamma}}\big(n_{\gamma}+1)+\frac{1}{3}\frac{ B_{\beta}}{B_{rot}}\Big(L(L+1)-K^2\Big)\Bigg)
\label{eq29}
\end{equation}
The allowed bands are characterized by
\begin{align}
n_{\gamma}&=0, && K=0; \nonumber\\ n_{\gamma}&=1, && K=\pm2;\nonumber\\
n_{\gamma}&=2, && K=0,\pm4; \  ...
\label{eq30}
\end{align}
In the standard case of constant mass  where $B_{\gamma}=B_{\beta}=B_{rot}=1$ and $\hbar=1$, our
 formula Eq. (\ref{eq29}) matches up with Eq. (41) of Ref ~\cite{bonatsos2}. In \cite{bonatsos2}, the coefficient of  $\gamma^2$ in $u(\gamma)$ is equal to $(3c)^2$, compared to $(\beta_0^4C_{\gamma})$ Eq. \eqref{eq26} used in this work.

The eigenfunctions corresponding to eigenvalues (\ref{eq28}) are obtained in terms of confluent hypergeometric function,
 \begin{equation}
\Gamma_{n_{\gamma},K}(\gamma)=N_{n_{\gamma},K \ 1}F_1\Big(-\tilde n_{\gamma},1+\frac{|K|}{2},\frac{\gamma^2}{g}\Big)
\label{eq311}
\end{equation}
where $N_{n_{\gamma},K}$ is a normalization constant. According to the relation between hypergeometric functions and the Laguerre polynomials, the $\gamma$ angular wave functions for axially symmetric prolate deformed nuclei can be written as:
\begin{equation}
\eta_{n_{\gamma},K}=N_{n_{\gamma},K}\ \gamma^{|K/2|}\ e^{-\frac{\gamma^2}{2g}}L_{\tilde n_{\gamma}}^{|K/2|}\Big(\frac{\gamma^2}{g}\Big)
\label{eq31}
\end{equation}
where $L_{\tilde n_{\gamma}}^{K/2}$ represents the Laguerre polynomial and $N_{\gamma}$ the normalization constant, determined from the normalization condition
\begin{equation}
\int_{0}^{\pi/3}\eta^2_{n_{\gamma},K}(\gamma)|\sin3\gamma|d\gamma=1
\label{eq321}
\end{equation}
In the case of small $\gamma$ vibration, we can write $|\sin3\gamma| \simeq |3\gamma|$, then the integral Eq. (\ref{eq321}) is easily calculated by using Eq. (8.980) of ~\cite{Gradshteyn}. This leads to
\begin{equation}
N_{n_{\gamma},K}=\left[\frac{2}{3}g^{-1-|K/2|}\frac{\tilde n_{\gamma}!}{\Gamma(\tilde n_{\gamma}+|K/2|+1)}\right]^{1/2}
\label{eq322}
\end{equation}
The normalization constants for the $(n_{\gamma},K)=(0,0)$ and $(n_{\gamma},K)=(1,2)$ states are found to be $N^2_{0,0}=\frac{2}{3g}$, $N^2_{1,2}=\frac{2}{3g^2}$ respectively, then $\frac{N^2_{0,0}}{N^2_{1,2}}=g$. This result will be used to calculate the $B(E2)$ values in $\gamma \longrightarrow$ ground and $\gamma \longrightarrow \beta$ transitions ($\Delta K=2$).
\section{\label{sec:level1} the radial schr\"{o}dinger equation}
The $\beta$-vibrations  states of deformed nuclei with mass parameter are determined by the solution of  the radial Schr\"{o}dinger equation
\begin{multline}
 \frac{\hbar^2}{2}\Bigg[\frac{1}{B_{\beta}}f^2F'' +\Big(\frac{1}{B_{\beta}}+\frac{1}{B_{\gamma}}\Big)\Big(ff'+\frac{2f^2}{\beta}\Big)F'\\+\Big(\frac{1}{B_{\beta}}+\frac{1}{B_{\gamma}}\Big)\Big(\frac{(f')^2}{8}+\frac{ff''}{4}+\frac{ff'}{\beta}\Big)F\Bigg]\\-\frac{f^2}{2\beta^2}\Lambda F+EF-V_{eff}F=0
   \label{eq32}
\end{multline}
with
\begin{multline}
 V_{eff}=V+\frac{\hbar^2}{2}\Big(\frac{1}{B_{\beta}}+\frac{1}{B_{\gamma}}\Big)\Bigg[\frac{1}{4}(1-\delta-\lambda)ff''\\+\frac{1}{2}\Big(\frac{1}{2}-\lambda\Big)\Big(\frac{1}{2}-\lambda\Big)(f')^2\Bigg]
   \label{eq33}
\end{multline}

Setting a standard transformation of the radial wave function
\begin{equation}
F_{n_{\beta}L}(\beta)=\beta^{-(1+B_{\beta}/B_{\gamma})}R_{n_{\beta}L}(\beta)
   \label{eq34}
\end{equation}
we get
\begin{multline}
-f^2R''-\Big(1+\frac{B_{\beta}}{B_{\gamma}}\Big)ff'R'-\Big(1+\frac{B_{\beta}}{B_{\gamma}}\Big)\Big(\frac{(f')^2}{8}+\frac{ff''}{4}\Big)R\\+2U_{eff}R=\frac{2B_{\beta}}{\hbar^2}ER    \label{eq35}
\end{multline}
where
\begin{multline}
U_{eff}=\frac{B_{\beta}}{\hbar^2}V_{eff}+\frac{1}{2}\frac{B_{\beta}}{B_{\gamma}}\Big(1+\frac{B_{\beta}}{B_{\gamma}}\Big)\frac{f^2+\beta ff'}{\beta^2}+\frac{B_{\beta}}{\hbar^2}\frac{f^2}{2\beta^2}\Lambda\label{eq36}
\end{multline}
In the frame without mass parameters, we can reduce the first three terms in the Eq. (\ref{eq39}) by $\big(\sqrt{f}\frac{d}{d\beta}\sqrt{f}\big)^2R$.

\section{\label{sec:level1} The effective potential and energy levels}
As in Ermamatov et al. work ~\cite{ermamatov}, we use in our  calculations the Davidson potential ~\cite{davidson}
\begin{equation}
V(\beta)=V_0\Bigg(\frac{\beta}{\beta_0}-\frac{\beta_0}{\beta}\Bigg)^2
   \label{eq37}
\end{equation}
where $V_0$ represent the depth of the minimum, located at $\beta_0$.

According to the specific form of the potential Eq. (\ref{eq37}), we are also going to consider for the deformation function the special form
\begin{align}
f(\beta)=1+a\beta^2, && a<<1
   \label{eq38}
\end{align}
Inserting these forms for the potential and the deformation
function in Eq. (\ref{eq36}) one gets
\begin{equation}
2U_{eff}=k_2\beta^2+k_0+\frac{k_{-2}}{\beta^2}
   \label{eq39}
\end{equation}
with
\begin{align}
k_2=&\frac{a^2}{2}\Bigg[  \Big(1+\frac{B_{\beta}}{B_{\gamma}}\Big)\Big(6\frac{B_{\beta}}{B_{\gamma}}+\big(1-2\delta\big)\big(1-2\lambda\big)     \nonumber \\
&+\big(1-\delta-\lambda\big)\Big)+2\frac{B_{\beta}}{\hbar^2}\Lambda \Bigg]+2\frac{g_{\beta}}{\beta_{0}^4} \nonumber&
\nonumber\\
k_0=&\frac{a}{2}\Bigg[\Big(1+\frac{B_{\beta}}{B_{\gamma}}\Big)\Big(8\frac{B_{\beta}}{B_{\gamma}}+\big(1-\delta-\lambda\big)\Big)+4\frac{B_{\beta}}{\hbar^2}\Lambda \Bigg]-4\frac{g_{\beta}}{\beta_{0}^2} \nonumber&
\nonumber\\
k_{-2}=&\frac{B_{\beta}}{B_{\gamma}}\Big(1+\frac{B_{\beta}}{B_{\gamma}}\Big)+\frac{B_{\beta}}{\hbar^2}\Lambda+2g_{\beta}&
  \label{eq40}
\end{align}
where $g_{\beta}=\frac{B_{\beta}V_0\beta_0^2}{\hbar^2}$.

To solve the radial equation Eq. (\ref{eq35}) through the asymptotic iteration method (AIM)~\cite{Iam}, one needs the following parametrization
\begin{align}
R_{n_{\beta}L}(y)=y^{\rho}(1+ay)^{\nu}\chi_{n_{\beta}L}(y), \     y=\beta^2
  \label{eq402}
\end{align}
where
\begin{align}
\rho&=\frac{1}{4}\big(1+\sqrt{1+4k_{-2}}\big)\nonumber\\
\nu&=-\frac{1}{2}\Bigg[\frac{B_{\beta}}{B_{\gamma}}+\Bigg(\Big(\frac{B_{\beta}}{B_{\gamma}}+\frac{1}{2}\Big)\Big(\frac{B_{\beta}}{B_{\gamma}}-1\Big)
+k_{-2}-\frac{k_0}{a}\nonumber \\&+\frac{k_2}{a^2}+\frac{2B_{\beta}}{a\hbar^2}E\Bigg)^{1/2}\Bigg]&
\label{eq47a}
\end{align}
For this form of the radial wave function, the  Eq. (\ref{eq35}) reads,
 \begin{align}
\chi''_{n_{\beta}}&(y)=-\Bigg[\frac{1+4\rho+ay(3+2\frac{B_{\beta}}{B_{\gamma}}+4\rho+4\nu)}{2y(1+ay)}\Bigg]\chi_{n_{\beta}}'(y)\nonumber\\
-a\Bigg[&\frac{2(\rho+\nu)(1+2\frac{B_{\beta}}{B_{\gamma}}+2\nu+2\rho)+1+\frac{B_{\beta}}{B_{\gamma}}-\frac{k_2}{a^2}}{4y(1+ay)}\Bigg]\chi_{n_{\beta}}(y)
  \label{eq403}
\end{align}
The first and the second terms in the right hand side of Eq. (\ref{eq403}) represent $\lambda_0$ and $s_0$ of Eq. (\ref{A.1}) respectively. After calculating $\lambda_n$ and $s_n$, by means of the recurrence relations of Eq. (\ref{A.4}), we get the generalized formula of the radial energy spectrum from the roots of the termination condition  of Eq. (\ref{A.3})
\begin{align}
E_{n_{\beta}n_{\gamma}LK}&=\frac{\hbar^2}{2B_{\beta}}\Big(k_0+\frac{a}{2}\big(2+\frac{B_{\beta}}{B_{\gamma}}+2p+2q+pq\big)\nonumber\\&+2a\big(2+p+q\big)n_{\beta}+4an_{\beta}^2\Big)
  \label{eq41}
\end{align}
where $n_{\beta}$ is the principal quantum number of $\beta$ vibrations, and
\begin{align}
q&\equiv q_{n_{\gamma}}(L,K)=\sqrt{1+4k_{-2}}\nonumber\\ p&\equiv p_{n_{\gamma}}(L,K)=\sqrt{4\frac{B_{\beta}}{B_{\gamma}}-3+4\frac{k_{2}}{a^2}}
  \label{eq42}
\end{align}

The quantities $k_2$, $k_0$, $k_{-2}$ are given by Eq. (\ref{eq40}), where $\Lambda$ is the eigenvalue of the $\gamma-$vibrational part of the Hamiltonian for axially symmetric prolate deformed nuclei.
In the numerical results part of the paper, the energies are normalized to the first excited state. So, the results depend on six parameters $B_{\beta}/B_{\gamma}$, $B_{\gamma}/B_{rot}$, $g$, $g_{\beta}$, $a$ and $\beta_0$.

A few interesting low-lying bands  are classified by the quantum numbers $n_{\beta}$, $n_{\gamma}$ and $K$, such as the ground state band (g.s.) with   $n_{\beta}=0$, $n_{\gamma}=0$, $K=0$,  the $\beta-$band with $n_{\beta}=1$, $n_{\gamma}=0$, $K=0$, and  the $\gamma-$band with $n_{\beta}=0$, $n_{\gamma}=1$, $K=2$.
\subsection{Special case 1: Without mass coefficients}
If we assume $B_{\beta}=B_{\gamma}=B_{rot}=1$, one gets from Eq.(\ref{eq40})
\begin{align}
k_2&=a^2\Big[\big(1-\delta-\lambda\big)+\big(1-2\delta\big)\big(1-2\lambda\big)+6+\Lambda \Big]+2\frac{V_0}{\beta_0^2} \nonumber&
\nonumber\\
k_0&=a\Big[\big(1-\delta-\lambda\big)
+8+2\Lambda\Big]-4V_0\nonumber&\nonumber\\
k&_{-2}=2+\Lambda+2V_0\beta^2_0&
  \label{eq43}
\end{align}
Thus, the energy spectrum formula Eq. (\ref{eq41}) is identical to Eq. (82) of Ref.~\cite{bonatsos2} obtained by means of supersymmetric quantum mechanical method (SUSYQM) \cite{Cooper1,Cooper2}. The slight difference between our coefficients $k_2$, $k_0$ and $k_{-2}$ and those of Ref.~\cite{bonatsos2} comes from the adopted expression of Davidson potential.

\subsection{Special case 2: No dependence of the mass on the deformation}
If $a=0$, the dependence of the mass on the deformation is canceled, then one obtains from Eq. (\ref{eq40})
\begin{align}
k_2&=2\frac{g_{\beta}}{\beta_{0}^4} , \hspace{1cm} k_0=-4\frac{g_{\beta}}{\beta_{0}^2}  \nonumber \\
k_{-2}&=\frac{B_{\beta}}{B_{\gamma}}\Big(1+\frac{B_{\beta}}{B_{\gamma}}\Big)+\frac{B_{\beta}}{\hbar^2}\Lambda+2g_{\beta}
 \label{eq44}
 \end{align}
In this case, the energy spectrum becomes
\begin{align}
E_{n_{\beta}n_{\gamma}LK}=\frac{\hbar^2}{2B_{\beta}}\Big(k_0+\sqrt{4k_2}\big(1+2n_{\beta}+\frac{1}{2}q_{n_{\gamma}}(L,K)\big)\Big)
 \label{eq45}
 \end{align}
 For axially symmetric prolate deformed nuclei, the energy formula reads
 \begin{align}
E_{n_{\beta}n_{\gamma}LK}=\sqrt{2\frac{V_0^2}{g_{\beta}}}\Big(1+2n_{\beta}+\frac{1}{2}q_{n_{\gamma}}(L,K)-\sqrt{2g_{\beta}}\Big)
 \label{eq46}
 \end{align}
 with
  \begin{align}
\frac{1}{2}q_{n_{\gamma}}(L,K)=\sqrt{\frac{1}{4}+\frac{B_{\beta}}{B_{\gamma}}\Big(1+\frac{B_{\beta}}{B_{\gamma}}\Big)+\frac{B_{\beta}}{\hbar^2}\Lambda+2g_{\beta}}
 \label{eq47}
 \end{align}
 and
   \begin{align}
\frac{B_{\beta}}{\hbar^2}\Lambda=\frac{2}{g}\frac{B_{\beta}}{B_{\gamma}}\big(n_{\gamma}+1)+\frac{1}{3}\frac{ B_{\beta}}{B_{rot}}\Big(L(L+1)-K^2\Big)
 \label{eq48}
 \end{align}
 Note that Eq. (\ref{eq46}) represents the correct formula of the energy spectrum, compared to Eq.(11) given in Ref. \cite{ermamatov}, where the mass parameter term is missed in the analogue formula of Eq. \eqref{eq47}.

It is also worth to note that, in this case, Eq. \eqref{eq32} reduces to standard confluent hypergeometric equation which can be converted to Laguerre differential equation. The resolution of such a problem is carried out in Section VIII.b.

 \subsection{Special case 3: Standard case }
For $\gamma$-unstable nuclei, in the limit case of $a=0$ and $B_{\beta}=B_{\gamma}=B_{rot}$, our formula Eq. (\ref{eq41}) reduces to
 \begin{align}
E_{n_{\beta}L}=\sqrt{2\frac{V_0^2}{g_{\beta}}}\Big(1+2n_{\beta}+\sqrt{\frac{9}{4}+\Lambda+2g_{\beta}}\Big)-2V_0
 \label{eq49}
 \end{align}
 with
  \begin{align}
\Lambda=\tau(\tau+3)
 \label{eq50}
 \end{align}
 and $\tau=L/2$ is the seniority quantum number. This formula  is similar to the energy spectrum Eq. (80) in Ref. ~\cite{rowe}.
 \section{Excited-state wave functions}
 The used wave functions in our calculations  are given by
 \begin{equation}
 \Psi(\beta,\gamma,\theta_i)=\beta^{-1-\frac{B_{\beta}}{B_{\gamma}}}R_{n_{\beta},L}(\beta)\eta_{n_{\gamma},K}(\gamma)\mathcal{D}_{M,K}^L(\theta_i)
 \label{eq51}
 \end{equation}
 The radial function $R_{n_{\beta},L}(\beta)$ corresponds to the  $n^{th}$ eigenstate of Eq. (\ref{eq35}), $\eta_{n_{\gamma},K}(\gamma)$ is given by Eq. (\ref{eq31}) and the symmetries eigenfunctions of the angular momentum are
  \begin{equation}
\mathcal{D}_{M,K}^L(\theta_i)=\sqrt{\frac{2L+1}{16\pi^2(1+\delta_{K0})}}\Big(D^{L*}_{MK}+(-1)^LD^{L*}_{M-K}\Big)
 \label{eq52}
 \end{equation}
To get the radial eigenvectors $R_{n_{\beta},L}(\beta)$ of Eq.
(\ref{eq35}), we insert the expression of the energy spectrum Eq.
\eqref{eq41} into Eq. \eqref{eq47a}. Then, we get from Eq.\eqref{eq403} and Eq. \eqref{eq402}  :
 \begin{align}
 R_{_{n_{\beta},L}}(y)=y^{\frac{1}{4}(1+q)}(1+ay)^{-n_{\beta}-\frac{1}{2}(1+\frac{B_{\beta}}{B_{\gamma}})-\frac{1}{4}(q+p)}\chi_{_{n_{\beta},L}}(y)
 \label{eq53}
  \end{align}
where $q$ and $p$ are given in Eq. (\ref{eq42}).\\
After inserting Eq. (\ref{eq53}) into Eq. (\ref{eq35}), we obtain
 \begin{align}
\chi_{n_{\beta},L}''(y)&=-\Bigg(\frac{1+\frac{q}{2}+a(1-2n_{\beta}-\frac{p}{2})y}{y(1+ay)}\Bigg)\chi_{n_{\beta},L}'(y)\nonumber\\&-\Bigg(\frac{an_{\beta}(n_{\beta}+\frac{p}{2})}{y(1+ay)}\Bigg)\chi_{n_{\beta},L}(y)
 \label{eq54}
  \end{align}
 The excited state wave functions of this equation are obtained through Eq. (\ref{A.2})
  \begin{align}
\chi(y)=N_{_{n_{\beta},L}2}F_1(-n_{\beta},-n_{\beta}-\frac{p}{2};-2n_{\beta}-\frac{(q+p)}{2};1+ay)
 \label{eq55}
  \end{align}
  where $N_{n_{\beta},L}$ is a  normalization constant and $_2F_1$ are hyper-geometrical functions.
 Therefore, according to the relation between hyper-geometrical functions and the
generalized Jacobi polynomials, Eq. (4.22.1) of Ref.~\cite{Szego}, the radial wave function can be written as
  \begin{align}
R_{n_{\beta},L}(t)&=N_{n_{\beta},L}2^{-(1+\frac{B_{\beta}}{B_{\gamma}})/2-(q+p)/4}a^{-(1+q)/4}\nonumber\\
&(1-t)^{(1+2\frac{B_{\beta}}{B_{\gamma}}+p)/4}(1+t)^{(q+1)/4}P_{n_{\beta}}^{(q/2,p/2)}(t),\nonumber\\
t&=\frac{-1+ay}{1+ay}
 \label{eq56}
  \end{align}
To determine $N_{n_{\beta},L}$, we use the  usual orthogonality relation of
Jacobi polynomials Eq. (7.391.7) of Ref.~\cite{Gradshteyn}. This leads to
  \begin{align}
  &N_{n_{\beta},L}=\Big(2a^{q/2+1}n_{\beta}!\Big)^{\frac{1}{2}}\nonumber\\
 & \Bigg(\frac{\Gamma\big(n_{\beta}+\frac{q+p}{2}+1\big)\Gamma\big(2n_{\beta}+\frac{q+p}{2}+1+\frac{B_{\beta}}{B_{\gamma}}\big)}{\Gamma\big(n_{\beta}+\frac{q}{2}+1\big)\Gamma\big(n_{\beta}+\frac{B_{\beta}}{B_{\gamma}}+\frac{p}{2}\big)\Gamma\big(2n_{\beta}+\frac{q+p}{2}+1\big)} \Bigg)^{\frac{1}{2}}
 \label{eq561}
  \end{align}
\paragraph{}
 In the case where  $B_{\beta}=B_{\gamma}=B_{rot}=1$ and $\hbar=1$, the wave function Eq. (\ref{eq56}) and the normalization constant Eq. (\ref{eq561}) match up with Eq. (108) and Eq. (112) of Ref.~\cite{bonatsos2} respectively.
 \paragraph{}
 In the limit case $a\longrightarrow0$, no dependence of the mass on the deformation,
 the second-order differential equation Eq. \eqref{eq35} must have a solution of the form
 \begin{align}
 R_{n_{\beta},L}(\beta)=\beta^{\frac{1}{2}(1+q_{n_{\gamma}}(L,k))}e^{-b\beta^2}G_{n_{\beta},L}(\beta)
 \label{eq57}
\end{align}
where $b=\sqrt{\frac{g_{\beta}}{2\beta_0^4}}$. By using this radial function in  Eq. \eqref{eq35} and introducing a new variable $y=\beta^2$ , one can get
 \begin{align}
G_{n_{\beta},L}''(y)&=-\Big(\frac{1+\frac{q}{2}-2by}{y}\Big)G_{n_{\beta},L}'(y)-\frac{2bn_{\beta}}{y}G_{n_{\beta},L}(y)
 \label{eq58}
  \end{align}
 From Eq.\eqref{A.1} of IAM, one can define $\lambda_0(0)$ and $s_0(y)$. Then, $\lambda_n(y)$ and $s_n(y)$ are calculated by the recurrence relations given in Eq. \eqref{A.4} and the solution of this equation is found through Eq. \eqref{A.2}  to be
   \begin{align}
G_{n_{\beta},L}(y)=N_{n_{\beta},L}L_{n_{\beta}}^{\frac{1}{2}q_{n_{\gamma}}(L,k)}\Big(2by\Big)
 \label{eq59}
  \end{align}
  where $L$ denotes the Laguerre polynomials, $N_{n_{\beta},L}$ is a normalization coefficient determined from the normalization condition
    \begin{align}
\int_{0}^{\infty}\beta^{2(1+\frac{B_{\beta}}{B_{\gamma})}}F^2(\beta)d\beta=1
\label{eq60}
  \end{align}

  leading to
        \begin{align}
N_{n_{\beta},L}=\Bigg[2(2b)^{\frac{1}{2}{q_{n_{\gamma}}(L,k)+1}}\frac{n_{\beta}!}{\Gamma(n_{\beta}+\frac{1}{2}q_{n_{\gamma}}(L,k)+1)}\Bigg]^{1/2}
\label{eq62}
  \end{align}
  \section{B(E2) TRANSITION RATES }
The electric quadrupole operator for axially deformed nuclei around $\gamma=0$ is given by ~\cite{iachello2}
    \begin{align}
T_M^{(E2)}=t\beta\Big[\mathcal{D}_{M,0}^{(2)} \cos\gamma + \frac{1}{\sqrt{2}}\Big(\mathcal{D}_{M,2}^{(2)}+\mathcal{D}_{\mu,-2}^{(2)}\Big)  \sin\gamma \Big]
\label{eq63}
  \end{align}
  where $t$ is a scaling factor. The first term describes $\Delta K=0$ transitions and the second is for $\Delta K=2$ transitions.

 The $B(E2)$ transition rates from an initial to a final state are given by \cite{Edmonds}
      \begin{align}
B(E2; L_iK_i\longrightarrow L_fK_f)=\frac{5}{16\pi}\frac{|\langle L_fK_f\| T^{(E2)} \| L_iK_i \rangle | ^2}{2L_i+1},
\label{eq64}
  \end{align}
  and the reduced matrix element can be obtained by using  the Wigner-Eckrat theorem \cite{Edmonds}
       \begin{align}
\langle L_fM_f&K_f|T_M^{(E2)}|L_iM_iK_i\rangle \nonumber
\\=&\frac{(L_i2L_f|M_iMM_f)}{\sqrt{2L_f+1}}\langle L_fK_f\| T^{(E2)} \| L_iK_i \rangle
\label{eq65}
  \end{align}
  The final result ~\cite{Bijker} reads
         \begin{align}
B(&E2;n_{\beta}Ln_{\gamma}K\longrightarrow n'_{\beta}L'n'_{\gamma}K')\nonumber\\
=&\frac{5}{16\pi}\langle L,K,2,K'-K|L',K'\rangle^2 I^2_{n_{\beta}L,n'_{\beta}L'}C^2_{n_{\gamma}K,n'_{\gamma}K'}
\label{eq66}
  \end{align}
with
\begin{align}
I_{n_{\beta},L,n'_{\beta},L'}&=\int \beta F_{L,n_{\beta}}(\beta)F_{L',n'_{\beta}}(\beta)\beta^{-2-2\frac{B_{\beta}}{B_{\gamma}}} d\beta \nonumber\\
&=\int \beta R_{L,n_{\beta}}(\beta)R_{L',n'_{\beta}}(\beta)d\beta
\label{eq67}
  \end{align}
 $C_{_{n_{\gamma}K,n'_{\gamma}K'}}$ contains the integral over $\gamma$. For $\Delta K=0$ corresponding to transitions  ($g.s.\rightarrow g.s., \gamma\rightarrow\gamma, \beta\rightarrow\beta$ and $\beta\rightarrow g.s. $), the $\gamma-$integral part reduces to the orthonormality condition of the $\gamma$-wave functions : $C_{_{n_{\gamma}K,n'_{\gamma}K'}}=\delta_{_{n_{\gamma},n'_{\gamma}}}\delta_{_{K,K'}}$. While for $\Delta K=2$ corresponding to transitions ($\gamma\rightarrow g.s., \gamma\rightarrow\beta$), this
integral takes the form.
\begin{align}
C_{n_{\gamma}K,n'_{\gamma}K'}=\int \sin\gamma\eta_{n_{\gamma}K}\eta_{n'_{\gamma}K'}|\sin3\gamma|d\gamma
\label{eq68}
  \end{align}
In the next sections, all values of $B(E2)$ are calculated in units of $B(E2;2^+_{1}\longrightarrow 0^+_{1})$.

\section{NUMERICAL RESULTS AND discussion }

\begingroup
\begin{table*}
\caption{\label{tab:Table1}The values of free parameters used in the calculations}
\begin{ruledtabular}
\begin{tabular}{lllllllllllll}
nucleus  & & $g$ && $g_{\beta}$& & $B_{\beta}/B_{\gamma}$ && $B_{\beta}/B_{rot}$&& $g(B_{\beta}=B_{\gamma}=B_{rot})$&& $g_{\beta}(B_{\beta}=B_{\gamma}=B_{rot})$\\
\hline

$^{154}$Sm && 0.0187&&281.66&&1.36&&3.99&&0.0489&&0.357\\
$^{156}$Gd && 0.0252&&308.84&&1.53&&4.64&&0.0673&&0.884\\
$^{172}$Yb & &0.0064&&2469.46&&1.32&&11.14&&0.0453&&-1.909\\
$^{182}$W && 0.0249&&619.74&&2.01&&6.62&&0.0714&&0.512\\
\end{tabular}
\end{ruledtabular}

\end{table*}
\endgroup

Before starting any calculations of the energy spectra and transition rates for the axially symmetric prolate deformed nuclei $^{154}$Sm, $^{156}$Gd, $^{172}$Yb, and $^{182}$W which have been the object of Ermamatov et al. study  ~\cite{ermamatov} and before trying to improve them within DDMF, we have to reevaluate the parameters of the problem through the corrected formulas of these nuclear characteristics Eqs. (\ref{eq46},\ref{eq47}). For this purpose, we determine the free parameters $B_{\gamma}/B_{\beta},g,$ and $ g_{\beta}$ from experimental data of $E(2^+_{\gamma})/E(2^+_{1})$, $E(0^+_{\beta})/E(2^+_{1})$ and  $B(E2;2^+_{\gamma}\longrightarrow 0^+_{1}/B(E2;2^+_{1}\longrightarrow 0^+_{1})$ by solving a system of three nonlinear algebraic equations (Appendix B), while $B_{\beta}/B_{rot}$ is fixed to the value given in~\cite{jolos2}.
With the new parameters (Table \ref{tab:Table1}) we have calculated the correct values that Ermamatov et al. \cite{ermamatov} should obtain for the ratios $E(L^+_{g.s})/E(2^+_{g.s.})$ for the ground state band,  $E(L^+_{\beta})/E(2^+_{g.s})$ for $\beta$-band and  $E(L^+_{\gamma})/E(2^+_{g.s.})$ for the $\gamma$-band. Here $E(L^+_i)$ ($i=g.s., \beta, \gamma$) is the energy of the level characterized by the angular momentum $L^+$ in the band $i$ and $E(2^+_{g.s.})$ the energy of the first excited level of the ground state band.
As a qualitative test of agreement between the theoretical results and the experimental data, we evaluated the $rms$ differences given by
\begin{equation}
\sigma=\sqrt{\frac{\Sigma^{n}_{i=1}(E_i(exp)-E_i(th))^2}{(n-1)E(2^+_1)^2}}
\label{eq69}
  \end{equation}
where $E_i(exp)$ is the experimental energy of the $i^{th}$ level, $E_i(th)$ the corresponding theoretical value, $n$ the maximum number of considered levels and $E(2^+_1)$ the head energy of the band under consideration.

In table \ref{tab:Table2}, we compare our results for $^{154}$Sm in the both cases $B_{\beta}\neq B_{\gamma}\neq B_{rot}$ (the third column with $a=0$) and  $B_{\beta}= B_{\gamma}= B_{rot}$ (the fifth column with $a=0$) with experimental data ~\cite{data} and the data from Ref. ~\cite{jolos2}. One can see that our results for $B_{\beta}\neq B_{\gamma}\neq B_{rot}$ agree with experimental data, particularly in $\beta$ and $\gamma$ bands ($\sigma<1$) but are slightly different from data of ~\cite{jolos2}. This slight discrepancy could be reduced in the frame of DDMF. While in the g.s. band the precision of our results ($\sigma>1$) is obviously affected by the energy value of the level $L=12$ which is nearly $10\%$ higher than the experimental one. From the same table, we can also see  that the obtained values in the case  $B_{\beta}\neq B_{\gamma}\neq B_{rot}$ are more precise ($\sigma_{total}<1$) than those for which $B_{\beta}= B_{\gamma}= B_{rot}$  ($\sigma_{total}>1$).
For $^{156}$Gd (table \ref{tab:Table3}) our results are relatively better in the $\gamma$ band for  $B_{\beta}\neq B_{\gamma}\neq B_{rot}$ but are globally more precise than for $B_{\beta}= B_{\gamma}= B_{rot}$. Moreover, our  energy spectrum for  $^{172}$Yb given in table  \ref{tab:Table4} well reproduce the standard ones, particularly in the g.s and $\gamma$ bands with $B_{\beta}\neq B_{\gamma}\neq B_{rot}$ unlike those of the case where these mass parameters are taken to be equal to one. On the other hand, our  results for the nucleus $^{182}$W (table \ref{tab:Table4}) are more accurate ($\sigma <1$) in the three bands with $B_{\beta}\neq B_{\gamma}\neq B_{rot}$ than in the case of $B_{\beta}= B_{\gamma}= B_{rot}$.

In order to improve the obtained numerical results, we recalculated the energy ratios in the framework of DDMF with the more elaborated formula given in Eq. \eqref{eq41}. Such a formula contains two supplementary parameters, namely : $a$ and $\beta_0$. The optimal values of  both parameters are evaluated through $rms$ fits of energy levels by making use of Eq. \eqref{eq69} for each band of each nucleus.

From tables \ref{tab:Table2}-\ref{tab:Table5} one can see that a fair enhancement of numerical results has been achieved within DDMF in  both cases : $B_{\beta}\neq B_{\gamma}\neq B_{rot}$ and $B_{\beta}= B_{\gamma}= B_{rot}$.

Indeed, from the numerical calculations for nuclei  $^{154}$Sm, $^{156}$Gd, $^{172}$Yb, and $^{182}$W, we remark that the precision in the case of $B_{\beta}\neq B_{\gamma}\neq B_{rot}$ increases with the mass number.

Similarly, we have also calculated transition rates $B(E2;L^+_{g.s.}+2\longrightarrow L^+_{g.s.})$, $B(E2;L^+_{\beta}\longrightarrow L^+_{g.s.})$ and $B(E2;L^+_{\gamma}\longrightarrow L^+_{g.s.})$  in units of  $B(E2;2^+_{g.s.}\longrightarrow 0^+_{g.s.})$ for the same nuclei in  both cases : $B_{\beta}\neq B_{\gamma}\neq B_{rot}$  and $B_{\beta}= B_{\gamma}= B_{rot}$ within and out DDMF. Within DDMF, we have used the same optimal values of the two parameters $a$ and $\beta_0$ previously obtained  for the energy ratios.

Then in tables \ref{tab:Table6}-\ref{tab:Table9}, it is clearly shown that our results in the case of $B_{\beta}\neq B_{\gamma}\neq B_{rot}$ are better than those with $B_{\beta}= B_{\gamma}= B_{rot}$. Indeed, in the $\beta$ band and in the case of different mass coefficients with $a=0$, the mean difference between the theoretical value of transition rate and the experimental one corresponding to transition $2^+_{\beta} \longrightarrow 0^+_{g.s.}$ is about 1.8, while in the case of equal mass parameters, it is equal to 20. Likewise, in the transition $2^+_{\beta} \longrightarrow 4^+_{g.s.}$, the mean difference between the theory and the experiment in the first case is about 15.9 and in the second one it is equal to 152. For $a\neq0$, in the case of different mass coefficients, for the same transitions $2^+_{\beta} \longrightarrow 0^+_{g.s.}$ and $2^+_{\beta} \longrightarrow 0^+_{g.s.}$, the mean difference value is about 1.7 and 14, respectively, while in the case of equal mass coefficients it is equal to 20.3 and 152.5, respectively. Such a fact can also be seen in the $\gamma$ band. We underline here that, in the case of equal mass parameters, the obtained results reproduce those of Bonatsos et al. \cite{bonatsos2}. The slight difference between them came from the fact that the Bonatsos et al. fitting calculations have been carried on a given number of levels which is different from the number we considered in our calculations. This is a further proof that our formulas given in Eq. \eqref{eq41} and Eq. \eqref{eq56} respectively for the energy and the wave functions are more accurate than those erroneously derived by Ermamatov et al. \cite{ermamatov}. Moreover, this comparison corroborates the fact that the mass parameter should be taken into account  in such calculations.
As it has been mentioned in the introduction, the Bohr Hamiltonian is a quite competitive method in respect to other methods like IBM-1 \cite{iachello87}. To make a simple comparison  between them, we give in tables \ref{tab:Table10}-\ref{tab:Table11}  our obtained results compared with the available IBM-1 data.

\begin{table*}[H]
\caption{\label{tab:Table2}The comparison of the theoretical predictions  of energy levels Eq. \eqref{eq41} of the ground state band, the $\beta$ and $\gamma$ bands normalized to the energy  of the first excited state $E(2_{g.s.}^+)$ using the parameters given in Table \ref{tab:Table1} for  $^{154}$Sm for  this work with those from Ref.~\cite{jolos2} and experimental values taken from Ref. \cite{data}.  $\beta_0$ and $a$   indicate the position of the minimum of Davidson potential Eq. \eqref{eq37} and the deformation dependence of the mass Eq. \eqref{eq38}  respectively, while $\sigma$ is the quality measure Eq. \eqref{eq69}.}
\begin{center}

\begin{tabular}{lllllllllllll}
\hline
\hline
 &&&\multicolumn {3}{c}{ $B_{\beta}\neq B_{\gamma}\neq B_{rot}$}&&&&\multicolumn {3}{c}{ $B_{\beta}= B_{\gamma}=B_{rot}$}\\
 \cline {4 -6}\cline {10 -12}\\
L&exp& &$a=0$ &&DDM&&&& $a=0$ &&DDM&Ref.~\cite{jolos2}\\
\hline
g.s.&\\
4&3.26&&3.31&&3.31&&&&3.25&&3.27&3.28\\
6&6.63&&6.89&&6.89&&&&6.59&&6.68&6.76\\
8&11.01&&11.65&&11.65&&&&10.82&&11.07&11.28\\
10&16.26&&17.52&&17.52&&&&15.75&&16.29&16.65\\
12&22.27&&24.41&&24.41&&&&21.22&&22.24&22.68\\
\hline
$\sigma$&&&1.289&&1.289 &&&&0.592&&0.043&0.320\\

$a$&&&&&0.0000&&&&&&0.0483&\\
$\beta_0$&&&&&22.41&&&&&&0.54&\\

\hline
 $\beta_1$&\\

0&13.40&&13.40&&13.03&&&&13.40&&13.13&13.40\\
2&14.37&&14.40&&14.03&&&&14.40&&14.14&14.40\\
4&16.32&&16.71&&16.35&&&&16.65&&16.41&16.68\\
6&19.23&&20.29&&19.94&&&&19.99&&19.78&20.16\\
8&&&25.05&&24.73&&&&24.22&&24.08&24.68\\
10&&&30.93&&30.64&&&&29.15&&29.11&30.05\\
12&&&37.81&&37.58&&&&34.62&&34.73&36.08\\
\hline
$\sigma$&&&0.651&&0.501&&&&0.479&&0.384&0.576\\
$a$&&&&&0.0335&&&&&&0.0039&\\
$\beta_0$&&&&&0.88&&&&&&0.88&\\
\hline
 $\gamma_1$&\\
2&17.56&&17.56&&18.01&&&&17.56&&18.67&17.56\\
3&18.77&&18.47&&18.97&&&&18.28&&19.49&16.56\\

4&20.30&&19.68&&20.26&&&&19.23&&20.57&19.87\\
5&22.01&&21.18&&21.86&&&&20.39&&21.91&21.48\\

6&23.73&&22.96&&23.78&&&&21.77&&23.50&23.38\\
7&26.27&&25.03&&26.01&&&&23.33&&25.33&25.53\\

8&&&27.36&&28.56&&&&25.08&&27.40&27.93\\
9&&&29.95&&31.41&&&&26.99&&29.70&30.55\\

10&&&32.79&&34.58&&&&29.05&&32.21&33.36\\
11&&&35.88&&38.06&&&&31.26&&34.94&36.33\\

12&&&39.19&&41.84&&&&33.60&&37.87&39.44\\
13&&&42.73&&45.93&&&&36.06&&40.99&42.65\\
\hline
$\sigma$&&&0.812&&0.260&&&&1.817&&0.743&1.097\\
$a$&&&&&0.0199&&&&&&0.0086&\\
$\beta_0$&&&&&11.53&&&&&&1.67&\\
\hline
$\sigma_{total}$&&&0.895&&0.874&&&&1.153&&0.754&0.728\\
$a$&&&&&0.0219&&&&&&0.0054&\\
$\beta_0$&&&&&1.07&&&&&&1.60&\\
\hline
\hline

\end{tabular}
\end{center}
\end{table*}

\begin{table*}
\caption{\label{tab:Table3}The comparison of the theoretical predictions  of energy levels Eq. \eqref{eq41} of the ground state band, the $\beta$ and $\gamma$ bands normalized to the energy  of the first excited state $E(2_{g.s.}^+)$ using the parameters given in Table \ref{tab:Table1} for  $^{156}$Gd for  this work with those from Ref.~\cite{jolos2} and experimental values taken from Ref. \cite{data}.  $\beta_0$ and $a$   indicate the position of the minimum of Davidson potential Eq. \eqref{eq37} and the deformation dependence of the mass Eq. \eqref{eq38}  respectively, while $\sigma$ is the quality measure Eq. \eqref{eq69}.}
\begin{center}
\begin{tabular}{lllllllllllll}
\hline
\hline
 &&&\multicolumn {3}{c}{ $B_{\beta}\neq B_{\gamma}\neq B_{rot}$}&&&&\multicolumn {3}{c}{ $B_{\beta}= B_{\gamma}=B_{rot}$}\\
 \cline {4 -6}\cline {10 -12}\\
L&exp& &$a=0$ &&DDM&&&& $a=0$ &&DDM&Ref.~\cite{jolos2}\\
\hline
g.s. &\\
4&3.24&&3.31&&3.30&&&&3.23&&3.25&3.29\\
6&6.57&&6.87&&6.82&&&&6.49&&6.60&6.76\\
8&10.84&&11.62&&11.50&&&&10.55&&10.87&11.22\\
10&15.91&&17.44&&17.30&&&&15.22&&15.92&16.51\\
12&21.63&&24.25&&24.18&&&&20.34&&21.62&22.38\\
\hline
$\sigma$&&&1.575&&1.492&&&&0.747&&0.025&0.470\\
$a$&&&&&0.0600&&&&&&0.0217&\\
$\beta_0$&&&&&60.00&&&&&&1.11&\\
\hline
$\beta_1$ &\\

0&11.79&&11.79&&9.93&&&&11.79&&11.79&11.79\\
2&12.69&&12.79&&10.94&&&&12.79&&12.79&12.79\\
4&14.68&&15.10&&13.29&&&&15.02&&15.02&15.08\\
6&17.30&&18.66&&16.93&&&&18.28&&18.28&18.54\\
8&20.76&&23.41&&21.82&&&&22.34&&22.34&23.01\\
10&24.94&&29.23&&27.89&&&&27.01&&27.01&28.30\\
12&30.43&&36.04&&35.09&&&&32.13&&32.13&34.17\\
\hline
$\sigma$&&&3.135&&2.585&&&&1.339&&1.339&2.311\\
$a$&&&&&0.0230&&&&&&0.0000&\\
$\beta_0$&&&&&2.77&&&&&&43.81&\\
\hline

 $\gamma_1$ &\\
2&12.97&&12.97&&12.97&&&&12.97&&13.75&12.97\\
3&14.02&&13.90&&13.90&&&&13.70&&14.58&13.96\\

4&15.22&&15.12&&15.12&&&&14.66&&15.67&15.27\\
5&16.93&&16.64&&16.64&&&&15.83&&17.02&16.88\\

6&18.47&&18.45&&18.45&&&&17.20&&18.61&18.76\\
7&20.79&&20.54&&20.54&&&&18.75&&20.44&20.90\\

8&22.60&&22.90&&22.90&&&&20.47&&22.50&23.27\\
9&25.28&&25.51&&25.51&&&&22.34&&24.78&25.85\\

10&27.44&&28.38&&28.38&&&&24.35&&27.28&28.60\\
11&30.19&&31.48&&31.48&&&&26.49&&29.97&31.50\\

12&32.84&&34.80&&34.80&&&&28.74&&32.86&34.52\\
13&35.67&&38.34&&38.34&&&&31.08&&35.94&37.63\\
\hline
$\sigma$&&&1.122&&1.122&&&&2.725&&0.390&0.982\\
$a$&&&&&0.0000&&&&&&0.0219&\\
$\beta_0$&&&&&32.35&&&&&&1.42&\\
\hline
$\sigma_{total}$&&&1.897&&1.866&&&&2.029&&1.008&1.379\\
$a$&&&&&0.0499&&&&&&0.0582&\\
$\beta_0$&&&&&0.98&&&&&&0.80&\\
\hline
\hline
\end{tabular}
\end{center}

\end{table*}



\begin{table*}
\caption{\label{tab:Table4}The comparison of the theoretical predictions  of energy levels Eq. \eqref{eq41} of the ground state band, the $\beta$ and $\gamma$ bands normalized to the energy  of the first excited state $E(2_{g.s.}^+)$ using the parameters given in Table \ref{tab:Table1} for  $^{172}$Yb  for  this work with those from Ref.~\cite{jolos2} and experimental values taken from Ref. \cite{data}.  $\beta_0$ and $a$   indicate the position of the minimum of Davidson potential Eq. \eqref{eq37} and the deformation dependence of the mass Eq. \eqref{eq38}  respectively, while $\sigma$ is the quality measure Eq. \eqref{eq69}. }
\begin{center}
\begin{tabular}{lllllllllllll}
\hline
\hline
 &&&\multicolumn {3}{c}{ $B_{\beta}\neq B_{\gamma}\neq B_{rot}$}&&&&\multicolumn {3}{c}{ $B_{\beta}= B_{\gamma}=B_{rot}$}\\
 \cline {4 -6}\cline {10 -12}\\
L&exp& &$a=0$ &&DDM&&&& $a=0$ &&DDM&Ref.~\cite{jolos2}\\
\hline
g.s.&\\
4&3.29&&3.33&&3.30&&&&3.25&&3.28&3.32\\
6&6.84&&6.96&&6.84&&&&6.58&&6.78&6.91\\
8&11.54&&11.87&&11.54&&&&10.79&&11.45&11.71\\
10&17.34&&18.02&&17.33&&&&15.69&&17.26&17.65\\
12&24.14&&25.36&&24.15&&&&21.12&&24.21&24.64\\
\hline
$\sigma$&&&0.719&&0.087&&&&1.766&&0.076&0.309\\
$a$&&&&&0.0036&&&&&&0.0800&\\
$\beta_0$&&&&&40.09&&&&&&1.98&\\
\hline
 $\beta_1$&\\

0&13.20&&13.20&&11.68&&&&13.20&&13.20&13.20\\
2&14.15&&14.20&&12.69&&&&14.20&&14.20&14.20\\
4&16.34&&16.53&&15.02&&&&16.45&&16.45&16.52\\
6&19.53&&20.16&&18.67&&&&19.78&&19.78&20.11\\
8&23.54&&25.07&&23.62&&&&23.99&&23.99&24.93\\
10&28.10&&31.22&&29.83&&&&28.89&&28.89&30.89\\
12&33.11&&38.56&&37.27&&&&34.32&&34.32&37.91\\
\hline
$\sigma$&&&4.593&&2.135&&&&0.628&&0.628&2.311\\
$a$&&&&&0.0418&&&&&&0.0000&\\
$\beta_0$&&&&&1.74&&&&&&1.60&\\
\hline
 $\gamma_1$&\\
2&18.63&&18.63&&18.71&&&&18.63&&19.08&18.63\\
3&19.68&&19.59&&19.69&&&&19.33&&19.89&19.63\\

4&21.06&&20.87&&20.98&&&&20.26&&20.95&20.95\\
5&22.60&&22.47&&22.60&&&&21.39&&22.28&22.60\\

6&&&24.38&&24.54&&&&22.73&&23.87&24.56\\
7&&&26.60&&26.80&&&&24.26&&25.73&26.83\\

8&&&29.12&&29.37&&&&25.96&&27.84&29.39\\
9&&&31.95&&32.26&&&&27.83&&30.20&32.24\\

10&&&35.07&&35.46&&&&29.85&&32.83&35.37\\
11&&&38.48&&38.97&&&&32.01&&35.70&38.76\\

12&&&42.17&&42.78&&&&34.29&&38.83&42.20\\
13&&&46.15&&47.91&&&&36.70&&42.22&46.28\\
\hline
$\sigma$&&&0.121&&0.065&&&&0.862&&0.347&0.070\\
$a$&&&&&0.0075&&&&&&0.0600&\\
$\beta_0$&&&&&17.10&&&&&&2.19&\\
\hline
$\sigma_{total}$&&&1.719&&1.413&&&&1.067&&3.458&1.495\\
$a$&&&&&0.0010&&&&&&0.0100&\\
$\beta_0$&&&&&11.12&&&&&&90.01&\\
\hline
\hline

\end{tabular}
\end{center}

\end{table*}

\begin{table*}
\caption{\label{tab:Table5}The comparison of the theoretical predictions  of energy levels Eq. \eqref{eq41} of the ground state band, the $\beta$ and $\gamma$ bands normalized to the energy  of the first excited state $E(2_{g.s.}^+)$ using the parameters given in Table \ref{tab:Table1} for  $^{182}$W  for  this work with those from Ref.~\cite{jolos2} and experimental values taken from Ref. \cite{data}.  $\beta_0$ and $a$   indicate the position of the minimum of Davidson potential Eq. \eqref{eq37} and the deformation dependence of the mass Eq. \eqref{eq38}  respectively, while $\sigma$ is the quality measure Eq. \eqref{eq69}.}
\begin{center}
\begin{tabular}{lllllllllllll}
\hline
\hline
 &&&\multicolumn {3}{c}{ $B_{\beta}\neq B_{\gamma}\neq B_{rot}$}&&&&\multicolumn {3}{c}{ $B_{\beta}= B_{\gamma}=B_{rot}$}\\
 \cline {4 -6}\cline {10 -12}\\
L&exp& &$a=0$ &&DDM&&&& $a=0$ &&DDM&Ref.~\cite{jolos2}\\
\hline

g.s.&\\
4&3.29&&3.32&&3.29&&&&3.22&&3.29&3.30\\
6&6.80&&6.91&&6.78&&&&6.45&&6.78&6.81\\
8&11.44&&11.71&&11.40&&&&10.47&&11.40&11.41\\
10&17.12&&17.65&&17.07&&&&15.05&&17.07&16.94\\
12&23.72&&24.64&&23.77&&&&20.07&&23.76&23.2\\
\hline
$\sigma$&&&0.548&&0.042&&&&2.161&&0.037&0.276\\
$a$&&&&&0.0335&&&&&&0.0470&\\
$\beta_0$&&&&&18.49&&&&&&1.25&\\
\hline
 $\beta_1$&\\

0&11.36&&11.36&&11.36&&&&11.36&&11.36&11.36\\
2&12.57&&12.36&&12.36&&&&12.36&&12.36&12.36\\
4&15.10&&14.68&&14.68&&&&14.58&&14.58&14.66\\
6&&&18.26&&18.26&&&&17.81&&17.81&18.17\\
8&&&23.07&&23.07&&&&21.83&&21.83&22.77\\
10&&&29.01&&29.01&&&&26.41&&26.41&28.30\\
12&&&36.00&&36.00&&&&31.43&&31.43&34.57\\
\hline
$\sigma$&&&0.335&&0.335&&&&0.395&&0.395&0.345\\
$a$&&&&&0.0000&&&&&&0.0000&\\
$\beta_0$&&&&&53.35&&&&&&52.55&\\
\hline
$\gamma_1$ &\\
2&12.21&&12.21&&12.42&&&&12.21&&12.76&12.21\\
3&13.31&&13.16&&13.19&&&&12.94&&13.55&13.21\\

4&14.43&&14.41&&14.46&&&&13.89&&14.60&14.52\\
5&16.24&&15.97&&16.03&&&&15.05&&15.90&16.14\\

6&17.70&&17.83&&17.90&&&&16.41&&17.42&18.05\\
7&&&19.71&&20.07&&&&17.95&&19.17&20.24\\

8&22.61&&22.41&&22.52&&&&19.65&&21.13&22.68\\
9&&&25.11&&25.26&&&&21.50&&23.28&25.35\\

10&&&28.08&&28.26&&&&23.48&&25.62&28.24\\
11&&&31.31&&31.53&&&&25.58&&28.14&31.31\\

12&&&34.78&&35.06&&&&27.79&&30.83&34.55\\
13&&&38.49&&38.83&&&&30.10&&33.68&37.92\\
\hline
$\sigma$&&&0.168&&0.158&&&&0.935&&0.380&0.194\\
$a$&&&&&0.0538&&&&&&0.0215&\\
$\beta_0$&&&&&0.95&&&&&&1.07&\\
\hline
$\sigma_{total}$&&&0.358&&0.357&&&&1.369&&1.019&0.240\\
$a$&&&&&0.0000&&&&&&0.0054&\\
$\beta_0$&&&&&50.35&&&&&&2.16&\\
\hline
\hline
\end{tabular}
\end{center}

\end{table*}


\begin{table*}
\caption{\label{tab:Table6}The comparison of the theoretical predictions of $B(E2)$  in units of  $B(E2;2^+_{g.s.}\longrightarrow 0^+_{g.s.})$ using the parameters given in Table  \ref{tab:Table1} for  $^{154}$Sm in this work with those from Ref.~\cite{jolos2} and experimental values.}
\begin{center}
\begin{tabular}{lllllllllll}
\hline
\hline
&&&\multicolumn {3}{c}{ $B_{\beta}\neq B_{\gamma}\neq B_{rot}$}&&\multicolumn {3}{c}{ $B_{\beta}= B_{\gamma}=B_{rot}$}\\
 \cline {4 -6}\cline {8 -10}\\
&exp&& $a=0$ &&DDM& &$a=0$ &&DDM&Ref.~\cite{jolos2}\\

\hline\\
$\frac{B(E2;L_{g.s.}+2\longrightarrow L_{g.s.})}{B(E2;2^+_{g.s.}\longrightarrow 0^+_{g.s.})}$&\\
\\
$ 4^+_{g.s.}\longrightarrow 2^+_{g.s.}$&1.40(5)&&1.44&&1.44&&1.47&&1.47&1.44\\
$ 6^+_{g.s.}\longrightarrow 4^+_{g.s.}$&1.67(7)&&1.61&&1.61&&1.70&&1.70&1.61\\
$ 8^+_{g.s.}\longrightarrow 6^+_{g.s.}$&1.83(11)&&1.72&&1.72&&1.90&&1.90&1.72\\
$10^+_{g.s.}\longrightarrow 8^+_{g.s.}$&1.81(11)&&1.81&&1.81&&2.10&&2.10&1.82\\
$12^+_{g.s.}\longrightarrow 10^+_{g.s.}$&&&1.90&&1.90&&2.30&&2.30&1.91\\
\hline\\
 $\frac{B(E2;L_{\beta}\longrightarrow L_{g.s.})}{B(E2;2^+_{g.s.}\longrightarrow 0^+_{g.s.})}\times 10^3$&\\
\\
$2^+_{\beta}\longrightarrow 0^+_{g.s.}$&5.4(13)&&6.5&&6.4&&23.3&&24.3&6.7\\
$4^+_{\beta}\longrightarrow 2^+_{g.s.}$&&&5.5&&5.5&&20.0&&21.3&5.6\\
$6^+_{\beta}\longrightarrow 4^+_{g.s.}$&&&3.1&&3.1&&12.4&&13.7&2.9\\
$2^+_{\beta}\longrightarrow 2^+_{g.s.}$&&&12.9&&11.9&&46.4&&47.9&13.3\\
$4^+_{\beta}\longrightarrow 4^+_{g.s.}$&&&11.7&&11.6&&42.3&&43.6&12.1\\
$6^+_{\beta}\longrightarrow 6^+_{g.s.}$&&&11.5&&11.4&&41.5&&42.9&11.9\\
$0^+_{\beta}\longrightarrow 2^+_{g.s.}$&&&59.7&&59.0&&216.5&&221.5&61.7\\
$2^+_{\beta}\longrightarrow 4^+_{g.s.}$&25(6)&&42.2&&35.6&&152.0&&154.4&43.8\\
$4^+_{\beta}\longrightarrow 6^+_{g.s.}$&&&48.5&&47.7&&169.9&&171.8&51.3\\
$6^+_{\beta}\longrightarrow 8^+_{g.s.}$&&&57.4&&56.2&&191.6&&193.1&62.8\\

\hline\\
 $\frac{B(E2;L_{\beta}\longrightarrow L_{g.s.})}{B(E2;2^+_{g.s.}\longrightarrow 0^+_{g.s.})}\times 10^3$&\\
 \\
$2^+_{\gamma}\longrightarrow 0^+_{g.s.}$&18.4(29)&&18.4&&14.8&&46.5&&48.7&18.4\\
$2^+_{\gamma}\longrightarrow 2^+_{g.s.}$&&&26.5&&21.3&&68.6&&71.4&26.2\\

$2^+_{\gamma}\longrightarrow 4^+_{g.s.}$&3.9(6)&&1.4&&1.1&&3.7&&3.8&1.3\\
$3^+_{\gamma}\longrightarrow 2^+_{g.s.}$&&&33.1&&29.3&&85.0&&88.8&32.8\\

$3^+_{\gamma}\longrightarrow 4^+_{g.s.}$&&&13.5&&11.8&&36.5&&37.6&13.0\\
$4^+_{\gamma}\longrightarrow 2^+_{g.s.}$&&&11.0&&9.7&&28.0&&29.4&11.0\\

$4^+_{\gamma}\longrightarrow 4^+_{g.s.}$&&&33.0&&28.7&&88.6&&91.7&32.1\\
$4^+_{\gamma}\longrightarrow 6^+_{g.s.}$&&&2.9&&2.4&&8.4&&7.8&2.7\\

$5^+_{\gamma}\longrightarrow 4^+_{g.s.}$&&&29.8&&25.8&&79.4&&82.65&29.3\\
$5^+_{\gamma}\longrightarrow 6^+_{g.s.}$&&&17.6&&14.8&&49.9&&51.0&16.5\\
\hline
\hline
\end{tabular}
\end{center}
\end{table*}


\begin{table*}
\caption{\label{tab:Table7}The comparison of the theoretical predictions of $B(E2)$  in units of  $B(E2;2^+_{g.s.}\longrightarrow 0^+_{g.s.})$ using the parameters given in Table  \ref{tab:Table1} for  $^{156}$Gd in this work with those from Ref.~\cite{jolos2} and experimental values.}
\begin{center}
\begin{tabular}{lllllllllll}
\hline
\hline
&&&\multicolumn {3}{c}{ $B_{\beta}\neq B_{\gamma}\neq B_{rot}$}&&\multicolumn {3}{c}{ $B_{\beta}= B_{\gamma}=B_{rot}$}\\
 \cline {4 -6}\cline {8 -10}\\
&exp&& $a=0$ &&DDM& &$a=0$ &&DDM&Ref.~\cite{jolos2}\\

\hline\\
$\frac{B(E2;L_{g.s.}+2\longrightarrow L_{g.s.})}{B(E2;2^+_{g.s.}\longrightarrow 0^+_{g.s.})}$&\\
\\
$ 4^+_{g.s.}\longrightarrow 2^+_{g.s.}$&1.41(5)&&1.44&&1.37&&1.48&&1.48&1.44\\
$ 6^+_{g.s.}\longrightarrow 4^+_{g.s.}$&1.58(6)&&1.61&&1.42&&1.73&&1.73&1.61\\
$ 8^+_{g.s.}\longrightarrow 6^+_{g.s.}$&1.71(10)&&1.72&&1.38&&1.96&&1.95&1.73\\
$10^+_{g.s.}\longrightarrow 8^+_{g.s.}$&1.68(9)&&1.82&&1.32&&2.19&&2.18&1.83\\
$12^+_{g.s.}\longrightarrow 10^+_{g.s.}$&1.60(16)&&1.91&&1.26&&2.43&&2.42&1.93\\
\hline\\
 $\frac{B(E2;L_{\beta}\longrightarrow L_{g.s.})}{B(E2;2^+_{g.s.}\longrightarrow 0^+_{g.s.})}\times 10^3$&\\
\\
$2^+_{\beta}\longrightarrow 0^+_{g.s.}$&3.4(3)&&6.1&&4.4&&24.7&&24.7&6.3\\
$4^+_{\beta}\longrightarrow 2^+_{g.s.}$&&&4.6&&3.9&&19.5&&19.5&4.7\\
$6^+_{\beta}\longrightarrow 4^+_{g.s.}$&&&2.1&&1.9&&11.0&&11.0&0.9\\
$2^+_{\beta}\longrightarrow 2^+_{g.s.}$&18(2)&&12.6&&4.2&&51.9&&51.9&13.0\\
$4^+_{\beta}\longrightarrow 4^+_{g.s.}$&&&11.5&&9.4&&47.3&&47.3&11.8\\
$6^+_{\beta}\longrightarrow 6^+_{g.s.}$&&&11.2&&3.6&&46.4&&46.4&11.6\\
$0^+_{\beta}\longrightarrow 2^+_{g.s.}$&&&60.5&&48.9&&251.1&&251.1&62.5\\
$2^+_{\beta}\longrightarrow 4^+_{g.s.}$&22(2)&&44.2&&1.9&&181.4&&181.4&46.0\\
$4^+_{\beta}\longrightarrow 6^+_{g.s.}$&&&52.0&&41.2&&204.9&&204.9&53.3\\
$6^+_{\beta}\longrightarrow 8^+_{g.s.}$&&&62.4&&48.9&&230.6&&230.6&68.9\\

\hline\\
 $\frac{B(E2;L_{\beta}\longrightarrow L_{g.s.})}{B(E2;2^+_{g.s.}\longrightarrow 0^+_{g.s.})}\times 10^3$&\\
 \\
$2^+_{\gamma}\longrightarrow 0^+_{g.s.}$&25.0(8)&&25.0&&25.0&&65.9&&68.3&25.0\\
$2^+_{\gamma}\longrightarrow 2^+_{g.s.}$&38.7(13)&&36.1&&36.1&&97.7&&100.6&35.5\\

$2^+_{\gamma}\longrightarrow 4^+_{g.s.}$&4.1(2)&&1.8&&1.8&&5.3&&5.3&1.8\\
$3^+_{\gamma}\longrightarrow 2^+_{g.s.}$&39.0(75)&&44.9&&44.9&&121.1&&125.2&44.6\\

$3^+_{\gamma}\longrightarrow 4^+_{g.s.}$&27.2(35)&&18.3&&18.3&&52.3&&53.3&17.7\\
$4^+_{\gamma}\longrightarrow 2^+_{g.s.}$&9.6(27)&&14.9&&14.9&&39.9&&41.5&14.9\\

$4^+_{\gamma}\longrightarrow 4^+_{g.s.}$&53.16(16)&&44.9&&44.9&&127.2&&130.3&43.6\\
$4^+_{\gamma}\longrightarrow 6^+_{g.s.}$&&&4.0&&4.0&&12.1&&11.1&3.7\\

$5^+_{\gamma}\longrightarrow 4^+_{g.s.}$&43(43)&&40.5&&40.5&&114.1&&117.6&39.8\\
$5^+_{\gamma}\longrightarrow 6^+_{g.s.}$&59(59)&&23.9&&23.9&&72.1&&72.8&22.4\\
\hline
\hline
\end{tabular}
\end{center}

\end{table*}


\begin{table*}
\caption{\label{tab:Table8}The comparison of the theoretical predictions of $B(E2)$  in units of  $B(E2;2^+_{g.s.}\longrightarrow 0^+_{g.s.})$ using the parameters given in Table  \ref{tab:Table1} for  $^{172}$Yb in this work with those from Ref.~\cite{jolos2} and experimental values.}
\begin{center}
\begin{tabular}{lllllllllll}
\hline
\hline
&&&\multicolumn {3}{c}{ $B_{\beta}\neq B_{\gamma}\neq B_{rot}$}&&\multicolumn {3}{c}{ $B_{\beta}= B_{\gamma}=B_{rot}$}\\
 \cline {4 -6}\cline {8 -10}\\
&exp&& $a=0$ &&DDM& &$a=0$ &&DDM&Ref.~\cite{jolos2}\\

\hline\\
$\frac{B(E2;L_{g.s.}+2\longrightarrow L_{g.s.})}{B(E2;2^+_{g.s.}\longrightarrow 0^+_{g.s.})}$&\\
\\
$ 4^+_{g.s.}\longrightarrow 2^+_{g.s.}$&1.42(10)&&1.43&&1.37&&1.47&&1.34&1.43\\
$ 6^+_{g.s.}\longrightarrow 4^+_{g.s.}$&1.51(7)&&1.59&&1.41&&1.70&&1.36&1.59\\
$ 8^+_{g.s.}\longrightarrow 6^+_{g.s.}$&1.89(19)&&1.67&&1.36&&1.90&&1.31&1.67\\
$10^+_{g.s.}\longrightarrow 8^+_{g.s.}$&1.77(11)&&1.74&&1.29&&2.11&&1.26&1.74\\
$12^+_{g.s.}\longrightarrow 10^+_{g.s.}$&&&1.79&&1.31&&2.32&&1.22&1.79\\
\hline\\
 $\frac{B(E2;L_{\beta}\longrightarrow L_{g.s.})}{B(E2;2^+_{g.s.}\longrightarrow 0^+_{g.s.})}\times 10^3$&\\
\\
$2^+_{\beta}\longrightarrow 0^+_{g.s.}$&1.1(1)&&2.4&&3.6&&23.5&&23.5&2.4\\
$4^+_{\beta}\longrightarrow 2^+_{g.s.}$&&&2.0&&2.1&&19.9&&19.9&2.0\\
$6^+_{\beta}\longrightarrow 4^+_{g.s.}$&&&1.1&&1.2&&12.3&&12.3&1.0\\
$2^+_{\beta}\longrightarrow 2^+_{g.s.}$&&&4.8&&4.6&&47.1&&47.1&4.17\\
$4^+_{\beta}\longrightarrow 4^+_{g.s.}$&&&4.4&&4.3&&42.8&&42.8&4.4\\
$6^+_{\beta}\longrightarrow 6^+_{g.s.}$&&&4.3&&4.2&&42.1&&42.1&4.3\\
$0^+_{\beta}\longrightarrow 2^+_{g.s.}$&&&22.3&&21.4&&220.3&&220.3&22.5\\
$2^+_{\beta}\longrightarrow 4^+_{g.s.}$&12(1)&&15.9&&6.3&&155.1&&155.1&16.0\\
$4^+_{\beta}\longrightarrow 6^+_{g.s.}$&&&18.4&&17.1&&173.7&&173.7&18.8\\
$6^+_{\beta}\longrightarrow 8^+_{g.s.}$&&&22.1&&20.3&&195.8&&195.8&22.8\\
\hline\\
 $\frac{B(E2;L_{\gamma}\longrightarrow L_{g.s.})}{B(E2;2^+_{g.s.}\longrightarrow 0^+_{g.s.})}\times 10^3$&\\
 \\

$2^+_{\gamma}\longrightarrow 0^+_{g.s.}$&6.3(5)&&6.3&&3.7&&42.3&&23.4&6.3\\
$2^+_{\gamma}\longrightarrow 2^+_{g.s.}$&&&9.0&&5.5&&62.6&&34.9&9.0\\

$2^+_{\gamma}\longrightarrow 4^+_{g.s.}$&0.60(5)&&0.5&&0.3&&3.4&&1.8&0.4\\
$3^+_{\gamma}\longrightarrow 2^+_{g.s.}$&&&11.3&&10.3&&77.5&&43.0&11.2\\

$3^+_{\gamma}\longrightarrow 4^+_{g.s.}$&&&4.6&&4.2&&33.4&&18.1&4.4\\
$4^+_{\gamma}\longrightarrow 2^+_{g.s.}$&33(24)&&3.8&&3.4&&25.5&&14.1&3.8\\

$4^+_{\gamma}\longrightarrow 4^+_{g.s.}$&&&11.1&&10.1&&81.1&&43.8&11.0\\
$4^+_{\gamma}\longrightarrow 6^+_{g.s.}$&&&1.0&&0.9&&7.7&&3.6&0.9\\

$5^+_{\gamma}\longrightarrow 4^+_{g.s.}$&&&10.1&&9.1&&72.5&&39.0&10.0\\
$5^+_{\gamma}\longrightarrow 6^+_{g.s.}$&&&5.8&&5.2&&45.9&&23.0&5.7\\
\hline
\hline
\end{tabular}
\end{center}

\end{table*}



\begin{table*}
\caption{\label{tab:Table9}The comparison of the theoretical predictions of $B(E2)$  in units of  $B(E2;2^+_{g.s.}\longrightarrow 0^+_{g.s.})$ using the parameters given in Table  \ref{tab:Table1} for  $^{182}$W in this work with those from Ref.~\cite{jolos2} and experimental values.}
\begin{center}
\begin{tabular}{lllllllllll}
\hline
\hline
&&&\multicolumn {3}{c}{ $B_{\beta}\neq B_{\gamma}\neq B_{rot}$}&&\multicolumn {3}{c}{ $B_{\beta}= B_{\gamma}=B_{rot}$}\\
 \cline {4 -6}\cline {8 -10}\\
&exp&& $a=0$ &&DDM& &$a=0$ &&DDM&Ref.~\cite{jolos2}\\
\hline\\
$\frac{B(E2;L_{g.s.}+2\longrightarrow L_{g.s.})}{B(E2;2^+_{g.s.}\longrightarrow 0^+_{g.s.})}$&\\
\\
$ 4^+_{g.s.}\longrightarrow 2^+_{g.s.}$&1.43(8)&&1.44&&1.36&&1.49&&1.48&1.44\\
$ 6^+_{g.s.}\longrightarrow 4^+_{g.s.}$&1.46(6)&&1.60&&1.40&&1.74&&1.72&1.60\\
$ 8^+_{g.s.}\longrightarrow 6^+_{g.s.}$&1.53(10)&&1.71&&1.35&&1.98&&1.92&1.71\\
$10^+_{g.s.}\longrightarrow 8^+_{g.s.}$&1.48(9)&&1.79&&1.28&&2.22&&2.11&1.80\\
$12^+_{g.s.}\longrightarrow 10^+_{g.s.}$&&&1.87&&1.23&&2.48&&2.30&1.88\\
\hline\\
 $\frac{B(E2;L_{\beta}\longrightarrow L_{g.s.})}{B(E2;2^+_{g.s.}\longrightarrow 0^+_{g.s.})}\times 10^3$&\\
\\
$2^+_{\beta}\longrightarrow 0^+_{g.s.}$&6.6(10)&&4.4&&4.4&&25.1&&25.1&5.2\\
$4^+_{\beta}\longrightarrow 2^+_{g.s.}$&&&3.2&&3.2&&19.2&&19.2&2.1\\
$6^+_{\beta}\longrightarrow 4^+_{g.s.}$&&&1.4&&1.4&&10.5&&10.5&0.1\\
$2^+_{\beta}\longrightarrow 2^+_{g.s.}$&4.6(6)&&9.3&&9.3&&53.6&&53.6&13.7\\
$4^+_{\beta}\longrightarrow 4^+_{g.s.}$&&&8.4&&8.4&&48.8&&48.8&12.5\\
$6^+_{\beta}\longrightarrow 6^+_{g.s.}$&&&8.3&&8.3&&47.9&&47.9&12.2\\
$0^+_{\beta}\longrightarrow 2^+_{g.s.}$&&&44.9&&44.9&&262.3&&262.3&77.0\\
$2^+_{\beta}\longrightarrow 4^+_{g.s.}$&13(1)&&33.1&&33.1&&191.1&&191.1&64.1\\
$4^+_{\beta}\longrightarrow 6^+_{g.s.}$&&&39.4&&39.4&&216.3&&216.3&81.3\\
$6^+_{\beta}\longrightarrow 8^+_{g.s.}$&&&47.81&&47.81&&243.2&&243.2&101.6\\

\hline\\
 $\frac{B(E2;L_{\gamma}\longrightarrow L_{g.s.})}{B(E2;2^+_{g.s.}\longrightarrow 0^+_{g.s.})}\times 10^3$&\\
 \\
$2^+_{\gamma}\longrightarrow 0^+_{g.s.}$&24.8(8)&&24.8&&24.8&&70.2&&72.1&24.8\\
$2^+_{\gamma}\longrightarrow 2^+_{g.s.}$&49.2(13)&&35.7&&35.7&&104.2&&106.5&35.3\\

$2^+_{\gamma}\longrightarrow 4^+_{g.s.}$&0.2(2)&&1.8&&1.8&&5.6&&5.7&1.8\\
$3^+_{\gamma}\longrightarrow 2^+_{g.s.}$&&&44.5&&44.6&&129.2&&132.5&44.2\\

$3^+_{\gamma}\longrightarrow 4^+_{g.s.}$&&&18.1&&18.1&&56.0&&56.8&17.6\\
$4^+_{\gamma}\longrightarrow 2^+_{g.s.}$&17.2(17)&&14.8&&14.8&&42.6&&43.8&14.8\\

$4^+_{\gamma}\longrightarrow 4^+_{g.s.}$&75.9(73)&&44.3&&44.3&&136.3&&138.7&43.3\\
$4^+_{\gamma}\longrightarrow 6^+_{g.s.}$&&&3.9&&3.8&&13.0&&11.8&3.7\\

$5^+_{\gamma}\longrightarrow 4^+_{g.s.}$&&&40.0&&40.1&&122.2&&125.0&39.5\\
$5^+_{\gamma}\longrightarrow 6^+_{g.s.}$&&&23.4&&23.4&&77.5&&78.1&22.3\\
\hline
\hline
\end{tabular}
\end{center}

\end{table*}


\begin{table*}[H]
\caption{\label{tab:Table10}The comparison of the theoretical predictions of $E(L^+_{i})$  ($i=g.s.,\beta,\gamma$-band) normalized to the energy  of the first excited state $E(2_{g.s.}^+)$ using the parameters given in Table \ref{tab:Table1}  for  $^{154}$Sm and $^{182}$W in this work with those from IBM-1 Ref.~\cite{sm,w182} and experimental values Ref.~\cite{data}.}
\begin{center}


\begin{tabular}{lllllllllllll}
\hline
\hline
 &&\multicolumn {2}{c}{ $^{154}Sm$}&&&&&&\multicolumn {2}{c}{ $^{182}W$}\\
 \cline {2 -6}\cline {9 -13}\\
L&exp& $a=0$ &DDM& IBM-1\cite{sm} &&&&exp&$a=0$ &DDM& IBM-1\cite{w182}&\\
\hline
g.s.&\\
4&3.26&3.31&3.31&3.19&&&&3.29&3.32&3.29&3.33&\\
6&6.63&6.89&6.89&7.33&&&&6.80&6.91&6.78&6.95&\\
8&11.01&11.65&11.65&12.44&&&&11.44&11.71&11.40&12.00&\\
10&&&&&&&&17.12&17.65&17.07&18.33&\\

\hline
$\sigma$&&0.490&0.490&1.127&&&&&0.350&0.039&0.775&\\
\hline
 $\beta_1$&\\

0&13.40&13.40&13.03&14.04&&&&11.36&11.36&11.36&11.41&\\
2&14.37&14.40&14.03&14.78&&&&12.57&12.36&12.36&11.46&\\
4&16.32&16.71&16.35&17.31&&&&15.10&14.68&14.68&13.81&\\
6&19.23&20.29&19.94&17.66&&&&&18.26&18.26&17.50&\\

\hline
$\sigma$&&0.651&0.501&1.158&&&&&0.332&0.332&1.204&\\
\hline
 $\gamma_1$&\\
2&17.56&17.56&18.01&18.53&&&&12.21&12.21&12.24&12.41\\
3&18.77&18.47&18.97&18.97&&&&13.31&13.16&13.19&12.47&\\
4&20.30&19.68&20.26&21.72&&&&14.43&14.41&14.46&14.94&\\
5&22.01&21.18&21.86&24.12&&&&16.24&15.97&16.03&15.48&\\
6&23.73&22.96&23.78&&&&&17.70&17.83&17.90&18.69&\\

\hline
$\sigma$&&0.623&0.298&1.576&&&&&0.168&0.158&0.801&\\
\hline
\hline
\end{tabular}
\end{center}
\end{table*}


\begin{table*}[H]
\caption{\label{tab:Table11}The comparison of the theoretical predictions of $B(E2)$  in units of  $B(E2;2^+_{g.s.}\longrightarrow 0^+_{g.s.})$ using the parameters given in Table  \ref{tab:Table1} for $^{154}$Sm and  $^{182}$W   in this work with those from IBM-1 Ref.~\cite{sm,w182} and experimental values Ref.~\cite{data}.}
\begin{center}


\begin{tabular}{lllllllllllll}
\hline
\hline
 &&\multicolumn {2}{c}{ $^{154}Sm$}&&&&&&\multicolumn {2}{c}{ $^{182}W$}\\
 \cline {2 -6}\cline {9 -13}\\
&exp& $a=0$ &DDM& IBM-1\cite{sm} &&&&exp&$a=0$ &DDM& IBM-1 \cite{w182} &\\
\hline
\\
$\frac{B(E2;L_{g.s.}+2\longrightarrow L_{g.s.})}{B(E2;2^+_{g.s.}\longrightarrow 0^+_{g.s.})}$&\\
\\
$ 4^+_{g.s.}\longrightarrow 2^+_{g.s.}$&1.40(5)&1.44&1.44&1.35&&&&1.43(8)&1.44&1.36&1.33&\\
$ 6^+_{g.s.}\longrightarrow 4^+_{g.s.}$&1.67(7)&1.61&1.61&1.53&&&&&&&&\\

\hline
\\
 $\frac{B(E2;L_{\beta}\longrightarrow L_{g.s.})}{B(E2;2^+_{g.s.}\longrightarrow 0^+_{g.s.})}\times 10^3$&\\
\\
$2^+_{\beta}\longrightarrow 0^+_{g.s.}$&5.4(13)&6.5&6.4&7.78&&&&6.6(10)&4.4&4.4&113.0&\\
$4^+_{\beta}\longrightarrow 2^+_{g.s.}$&&5.5&5.5&29.57&&&&&3.2&3.2&56.3&\\

$2^+_{\beta}\longrightarrow 2^+_{g.s.}$&&12.9&12.8&15.33&&&&4.6(6)&9.3&9.3&48.4&\\

$2^+_{\beta}\longrightarrow 4^+_{g.s.}$&25(6)&42.2&41.4&30.67&&&&13(1)&33.1&33.1&3.7&\\

\hline
\\
 $\frac{B(E2;L_{\beta}\longrightarrow L_{g.s.})}{B(E2;2^+_{g.s.}\longrightarrow 0^+_{g.s.})}\times 10^3$&\\
 \\
 $2^+_{\gamma}\longrightarrow 0^+_{g.s.}$&18.4(29)&18.4&16.5&16.43&&&&&&&&\\
$2^+_{\gamma}\longrightarrow 2^+_{g.s.}$&&26.5&23.6&24.10&&&&&&&&\\

$2^+_{\gamma}\longrightarrow 4^+_{g.s.}$&3.9(6)&1.4&1.2&0.99&&&&0.2(2)&1.8&1.8&153.6&\\

\hline
\hline
\end{tabular}
\end{center}
\end{table*}

\section{Conclusion }
In this paper we have revisited all calculations performed in a recent work \cite{ermamatov} based upon inaccurate formulas for the energy spectrum and transition rates for axially symmetric prolate nuclei. With the asymptotic iteration method we have derived the correct formulas for these nuclear observables. Also, we have extended our calculations into deformation dependent effective masses formalism in order to improve the numerical results. Moreover, we have shown the importance of the mass parameter to be introduced in numerical calculations unlike what it has been done by other authors who have neglected the important role played by this parameter in such calculations.
Through a comparison with IBM-1, the Bohr Hamiltonian with mass parameters has proved to be more accurate.
%
\appendix
\section{ Asymptotic Iteration Method (AIM)}
The asymptotic iteration method \cite{Iam} is proposed to solve the second-order homogeneous differential
equation of the form
\begin{equation}
y''(x)=\lambda_0(x)y'(x)+s_0(x)y(x)
\label{A.1}
\end{equation}
where the variables $\lambda_0$ and $s_0$ are sufficiently differentiable.\\
The differential equation (\ref{A.1}) has a general solution \cite{Iam}
\begin{align}
y(x)&=\exp\Big(-\int^{x}\alpha(x_1)dx_1\Big)\Big[C_2\nonumber \\+&C_1\int^{x}\exp\Big(\int^{x_1}[\lambda_0(x_2)+2\alpha(x_2)]dx_2
   \Big)dx_1\Big] \label{A.2}
\end{align}

If we have $n>1$, for sufficiently large n, $\alpha(x)$ values can be obtained
\begin{equation}
\frac{s_n(x)}{\lambda_n(x)}=\frac{s_{n-1}(x)}{\lambda_{n-1}(x)}=\alpha(x)
  \label{A.21}
\end{equation}
with the sequences
\begin{subequations}
  \begin{align}
  \lambda_n(x)=&\lambda'_{n-1}(x)+s_{n-1}(x)+\lambda_0(x)\lambda_{n-1}(x)    \\
   s_n(x)=&s'_{n-1}(x)+s_0(x)\lambda_{n-1}(x), && n=1,2,3....
  \end{align}
  \label{A.4}
  \end{subequations}
the energy eigenvalues are then computed by means of the following termination condition \cite{Iam}
\begin{equation}
\delta=s_n\lambda_{n-1}-\lambda_ns_{n-1}=0
  \label{A.3}
\end{equation}
\section{ Formulas used for the calculations of the $B(E2)$ }
In this appendix we present the expressions used for calculations of the  transition probabilities $B(E2)$ :
 \begin{align}
&  \frac{B(E2;L'^+_{g.s.}\longrightarrow L^+_{g.s.})}{B(E2;2^+_{g.s.}\longrightarrow 0^+_{g.s.})} \nonumber \\
 & =5(C^{L0}_{L'020})^2 \nonumber \\
 &\times\Bigg(\frac{\Gamma\big[0.5\big(q_0(L',0)+q_0(L,0)\big)+1.5\big]}{\Gamma\big[0.5\big(q_0(2,0)+q_0(0,0)\big)+1.5\big]}\Bigg)^2 \nonumber \\
 &\times\frac{\Gamma\big[q_0(2,0)+1\big] \Gamma\big[q_0(0,0)+1\big]}{\Gamma\big[q_0(L',0)+1\big] \Gamma\big[q_0(L,0)+1\big]}
   \label{B.1}
  \end{align}
\begin{align}
&  \frac{B(E2;L'^+_{\beta}\longrightarrow L^+_{g.s.})}{B(E2;2^+_{g.s.}\longrightarrow 0^+_{g.s.})} \nonumber \\
 & =\frac{5}{4}(C^{L0}_{L'020})^2 \nonumber \\
 &\times\Bigg(\frac{\Gamma\big[0.5\big(q_0(L',0)+q_0(L,0)\big)+1.5\big]}{\Gamma\big[0.5\big(q_0(2,0)+q_0(0,0)\big)+1.5\big]}\Bigg)^2 \nonumber \\
 &\times\frac{\Gamma\big[q_0(2,0)+1\big] \Gamma\big[q_0(0,0)+1\big]}{\Gamma\big[q_0(L',0)+1\big] \Gamma\big[q_0(L,0)+1\big]} \nonumber \\
 &\times \frac{(q_0(L',0)-q_0(L,0)-1)^2}{q_0(L',0)+1}
   \label{B.2}
  \end{align}
 \begin{align}
&  \frac{B(E2;L'^+_{\gamma.}\longrightarrow L^+_{g.s.})}{B(E2;2^+_{g.s.}\longrightarrow 0^+_{g.s.})} \nonumber \\
 & =5g(C^{L0}_{L'020})^2 \nonumber \\
 &\times\Bigg(\frac{\Gamma\big[0.5\big(q_0(L',0)+q_0(L,0)\big)+1.5\big]}{\Gamma\big[0.5\big(q_0(2,0)+q_0(0,0)\big)+1.5\big]}\Bigg)^2 \nonumber \\
 &\times\frac{\Gamma\big[q_0(2,0)+1\big] \Gamma\big[q_0(0,0)+1\big]}{\Gamma\big[q_0(L',0)+1\big] \Gamma\big[q_0(L,0)+1\big]}
   \label{B.3}
  \end{align}
where $C^{L0}_{L'020}$ is Clebsch-Gordan coefficients.
%



\begin{thebibliography}{}
\bibitem{bohr}
A. Bohr, Mat. Fys. Medd. K. Dan. Vidensk. Selsk. {\bf 26},
no. 14 (1952).
\bibitem{bohr2}
A. Bohr and B. R. Mottelson, {\it Nuclear Structure Vol. II:
Nuclear Deformations} (Benjamin, New York, 1975).
\bibitem{eisenberg}
J. M. Eisenberg and W. Greiner,{\it Nuclear Theory Vol.I:
Nuclear Models} (North-Holland, Amsterdam, 1975).
\bibitem{rin80}
P. Ring and P. Schuck, {\it The Nuclear Many-Body Problem} (Springer Verlag, New York, 1980).
\bibitem{qrpa11}
J. Terasaki and J. Engel, Phys. Rev. C {\bf84}, 014332 (2011).
\bibitem{iachello87}
F. Iachello and A. Arima, {\it The Interacting Boson Model}
(Cambridge University Press, Cambridge, 1987).

\bibitem{fort03}
L. Fortunato and A. Vitturi, J. Phys. G: Nucl. Part.
Phys. {\bf29}, 1341 (2003).
\bibitem{fort0430}
L. Fortunato and A. Vitturi, J. Phys. G: Nucl. Part.
Phys. {\bf30}, 627 (2004).
\bibitem{fot04}
L. Fortunato, Phys. Rev. C {\bf70}, 011302 (2004).
\bibitem{fortunato}
L. Fortunato, Eur. Phys. J. A {\bf26} (s01), 1 (2005).
\bibitem{bonat04}
D. Bonatsos, D. Lenis, N. Minkov, P. P. Raychev and P. A. Terziev, Phys. Rev. C {\bf69}, 014302 (2004).
\bibitem{bonat07}
D. Bonatsos, E. A. McCutchan, N. Minkov, R. F. Casten, P. Yotov, D. Lenis, D. Petrellis and I. Yigitoglu, Phys. Rev.C {\bf76}, 064312 (2007).
\bibitem{bonat11}
I. Yigitoglu and D. Bonatsos, Phys. Rev. C {\bf83}, 014303 (2011).
\bibitem{bonatsos2}
D. Bonatsos, P. E. Georgoudis, D. Lenis, N. Minkov and C. Quesne, Phys. Rev. C {\bf83}, 044321 (2011).
\bibitem{bona2013}
D. Bonatsos, P. E. Georgoudis, D. Lenis, N. Minkov, D. Petrellis and C. Quesne, Phys. Rev. C {\bf88},  034316 (2013).
\bibitem{jolos0}
R. V. Jolos and P. von Brentano, Phys. Rev. C {\bf 74}, 064307 (2006).
\bibitem{jolos}
R. V. Jolos and P. von Brentano, Phys. Rev. C {\bf 76}, 024309 (2007).
\bibitem{jolos2}
R. V. Jolos and P. von Brentano, Phys. Rev. C {\bf78}, 064309 (2008).
\bibitem{jolos3}
R. V. Jolos and P. von Brentano, Phys. Rev. C {\bf79}, 044310 (2009).
\bibitem{ermamatov}
M. J. Ermamatov and P. R. Fraser, Phys. Rev. C {\bf84}, 044321 (2011).
\bibitem{ermamatov2}
S. Sharipov and M. J. Ermamatov, Int. J. Mod. Phys. E {\bf12}, 41 (2003).
\bibitem{davidson}
P. M. Davidson, Proc. R. Soc. London Ser. A {\bf135}, 459 (1932).
\bibitem{Cooper1}F. Cooper, A. Khare and U. Sukhatme, Phys. Rep. {\bf251},
267 (1995).
\bibitem{Cooper2} F. Cooper, A. Khare and U. Sukhatme,{\it Supersymmetry
in Quantum Mechanics} (World Scientific, Singapore, 2001).
\bibitem{Iam}
H. Ciftci, R. L. Hall and N. Saad, J. Phys. A {\bf36}, 11807 (2003).
\bibitem{Iam1}
M. Chabab and M. Oulne, Int. Rev. Phys. {\bf4}, 331 (2010).
\bibitem{Iam2}
M. Chabab, R. Jourdani and M. Oulne, Int. J. Phys. Sci. {\bf7}, 1150 (2012).
\bibitem{Iam3}
M. Chabab, A. Lahbas and M. Oulne, Int. J. Mod. Phys. E {\bf21}, 10 (2012).
 \bibitem{roos83}
 O. von Roos, Phys. Rev. B  {\bf 27}, 7547 (1983)
  \bibitem{quesne2004}
C. Quesne and V. M. Tkachuk, J. Phys. A: Math. Gen.
 {\bf 37}, 4267 (2004).
\bibitem{wilet}
L. Wilets and M. Jean, Phys. Rev. {\bf 102}, 788 (1956).
\bibitem{iachello}
F. Iachello, Phys. Rev. Lett. {\bf87}, 052502 (2001).
\bibitem{rowe}
D. J. Rowe and C. Bahri, J. Phys. A {\bf31}, 4947 (1998).
\bibitem{Szego}
G. Szego, {\it Orthagonal Polynomials} (American Mathematical Society, New York, 1939).
\bibitem{iachello2}
F. Iachello, Phys. Rev. Lett. {\bf87}, 052502 (2001).
\bibitem{Edmonds}
A. R. Edmonds,{\it Angular Momentum in Quantum Mechanics} (Princeton University Press, Princeton, 1957).
\bibitem{Bijker}
R. Bijker, R. F. Casten, N. V. Zamfir and E. A. McCutchan,
Phys. Rev. C {\bf68}, 064304 (2003).
\bibitem{abramowitz}
M. Abramowitz and I. A. Stegun, {\it Handbook of Mathematical Functions} (Dover, New York, 1965).
\bibitem{Gradshteyn}
I. S. Gradshteyn and I. M. Ryzhik, {\it Table of Integral, Series, and Products }(Academic, New York, 1980).
\bibitem{data}
http://www.nndc.bnl.gov/nndc/ensdf/.
\bibitem{sm}
N. Abood Saad, A. Najim Laith and Kh. Jundi, EJAE. {\bf1}, 3 (2014).
\bibitem{w182}
A. A. Abojassem and F. A. AL Temame, J. Kufa Phys. {\bf3}, 2 (2011).

\end{thebibliography}
%

\end{document}